\documentclass[12pt]{iopart}
\usepackage{setstack}
\usepackage{iopams}
\usepackage{graphicx}
\usepackage{esint}
\usepackage{float}
\usepackage{units}
\usepackage[all]{xy}

\newtheorem{theo}{Theorem}
\newtheorem{defi}{Definition}

\newtheorem{lem}{Lemma}
\newtheorem{ex}{Example}
\newtheorem{fact}{Fact}

\begin{document}
\title[Discrete Morse functions for graph configuration spaces]{Discrete Morse functions for graph configuration spaces}
\author{A Sawicki$^{1,2}$}

\address{$^1$School of Mathematics, University of Bristol,
University Walk, Bristol BS8 1TW, UK}

\address{$^2$Center for Theoretical Physics, Polish Academy of Sciences, Al.
Lotnik\'ow 32/46, 02-668 Warszawa, Poland}

\eads{\mailto{Adam.Sawicki@bristol.ac.uk}}

\begin{abstract}
We present an alternative application of discrete Morse theory for two-particle graph configuration spaces. In contrast to previous constructions, which are based on discrete Morse vector fields, our approach is through Morse functions, which have a nice physical interpretation as two-body potentials constructed from one-body potentials.  We also give a brief introduction to discrete Morse theory. Our motivation comes from the problem of quantum statistics for particles on networks, for which generalised  versions of anyon statistics can appear.


\end{abstract}
\submitto{\JPA}
\maketitle

\section{Introduction}

In non-relativistic quantum mechanics a quantum system is described by a wavefunction which fulfills the Schr\"{o}dinger equation. Moreover, for a system of many identical particles the additional symmetrization (for bosons) or antisymmetrization (for fermions) of the wavefunction is imposed. Around a quarter-century ago Souriau \cite{Souriau70}, Leinaas and  Myrheim \cite{leinass}, and subsequently Wilczek \cite{wilczek}, noticed that for identical particles confined to two dimensions there are other possibilities, namely one can have exotic quantum statistics, aka anyons. Recently \cite{JHJKJR} Harrison, Keating and Robbins (see also \cite{BE}) discussed the quantum abelian statistics of two
indistinguishable spinless particles on a quantum graph. They found that in spite of the fact that the model is locally one-dimensional, anyon statistics are present. Moreover, they noticed that at least a priori there is a possibility of having more than one statistics phase. The analysis described in \cite{JHJKJR} is simplified by considering combinatorial, rather than metric, graphs i.e.~many-particle tight-binding models. It was shown in \cite{Abrams} that under some
further assumptions, which are specified in section \ref{sec:Quantum-statisctics-for-Cn}, many-particle combinatorial graphs have the same topological properties as their metric counterparts and hence combinatorial graphs are equivalent
to metric ones from the point of view of quantum statistics. As discussed in section \ref{sec:Quantum-statisctics-and-fundamental}, the first homology group of an appropriate configuration space is related to quantum statistics \cite{Dowker85,JHJKJR}.

Recently there has been significant progress in understanding topological properties of configuration spaces of many particles on metric graphs \cite{farley,kiko}. This  was enabled by the foundational development of discrete Morse theory by Forman during  the late 1990's \cite{forman}. This theory reduces the calculation of homology groups to an essentially combinatorial problem, namely the construction of certain discrete Morse functions, or equivalently discrete gradient vector fields. Using this idea Farley and Sabalka \cite{farley} gave a recipe for the construction of such a discrete gradient vector field \cite{farley} on many-particle graphs and classified the first homology groups for tree graphs.
In 2011 Ko and Park \cite{kiko} significantly extended these results to arbitrary graphs by incorporating graph-theoretic theorems concerning the decomposition of a graph into its two and three-connected components.

In the current paper we give an alternative application of discrete Morse theory for two-particle graph configuration spaces. In contrast to the construction given in \cite{farley}, which is based on discrete Morse vector fields, our approach is through discrete Morse functions. Our main goal is to provide an intuitive way of constructing a discrete Morse function and hence a discrete Morse gradient vector field. The central object of the construction is the `trial Morse' function. It may be understood as two-body potential constructed from one-body potential, a perspective which is perhaps more natural and intuitive from a physics point of view. Having a perfect Morse function $f_1$ on a graph $\Gamma$ we treat it as a one-body potential. The value of the trial Morse function at each point of a two-particle configuration space is the sum of the values of $f_{1}$ corresponding to the two particles positions in $\Gamma$. The trial Morse function is typically not a Morse function, i.e. it might not satisfy some of the relevant conditions. Nevertheless, we find that it is always possible to modify it and obtain a proper Morse function out of it. In fact, the trial Morse function is not `far' from being a Morse function and the number of cells at which it needs fixing is relatively small. Remarkably, this simple idea leads to similar results as those obtained in \cite{farley}. We demonstrate it in Section 5 by calculating two simple examples. We find that in both cases the trial Morse function has small defects which can be easily removed and a proper Morse function is obtained. The corresponding discrete Morse vector field is equivalent to the one stemming from the Farley and Sabalka method \cite{farley}. As is shown in Section 7, it is always possible to get rid of defects of the trial Morse function. The argument is rather technical. However, since the problem is of a certain combinatorial complexity we believe it cannot be easily simplified. We describe in details how the final result, i.e set of discrete Morse functions along with rules for identifying the critical cells and constructing the boundary map of the associated Morse complex, is built in stages from this simple idea. Our main purpose is hence to present an approach which we believe is conceptually simple and physically natural. It would be interesting to check if the presented constructions can give any simplification in understanding the results of \cite{kiko} but we do not pursue this here.

The paper is organized as follows. In section 2 we discuss the relation between quantum statistics and the first homology group of a configuration space. In section \ref{sec:Morse-theory-in} we give a brief introduction to discrete Morse theory. Then in sections \ref{sec:One-particle-graph} and~\ref{sec:Main-example}, for two examples we present a definition of a `trial' Morse function $\tilde{f}_2$ for two-particle graph configuration space. We notice that the trial Morse function typically does not satisfy the conditions required of a Morse function according to Forman's theory. Nevertheless, we show in Section 7 that with small modifications, which we explicitly identify, the trial Morse function can be transformed into a proper Morse function. Since the number of critical cells and hence the size of the associated Morse complex is small compared with the size of configuration space the calculation of homology groups are greatly simplified. The technical details of the proofs are given in the Appendix. In section 6 we discuss more specifically how the techniques of discrete Morse theory apply to the problem of quantum statistics on graphs.


\section{Quantum statistics and the fundamental group \label{sec:Quantum-statisctics-and-fundamental}}

Symmetrization (for bosons) and anti-symmetrization (for fermions)
of the Hilbert space of indistinguishable particles is typically introduced
as an additional postulate of non-relativistic quantum mechanics.
More precisely, for indistinguishable particles the Hilbert space of
a composite, $n$-partite system is not the tensor product of
the single-particle Hilbert space but rather,
\begin{enumerate}
\item the antisymmetric part of the tensor product, for fermions,

\item the symmetric part of the tensor product, for bosons.

\end{enumerate}
In terms of the  wave function in the position representation this translates
to
\begin{eqnarray*}
\Psi(x_{1},\ldots,x_{i},\ldots,x_{j},\ldots x_{n})=\Psi(x_{1},\ldots,x_{j},\ldots,x_{i},\ldots x_{n})\,\,\,\,\,\mathrm{for}\,\,\mathrm{bosons},\\
\Psi(x_{1},\ldots,x_{i},\ldots,x_{j},\ldots x_{n})=-\Psi(x_{1},\ldots,x_{j},\ldots,x_{i},\ldots x_{n})\,\,\,\,\,\mathrm{for}\,\,\mathrm{fermions},
\end{eqnarray*}
i.e., when two fermions are exchanged the sign of wave function changes
and for bosons it stays the same.

It was first noticed by Souriau \cite{Souriau70}, and subsequently by Leinaas and Myrheim \cite{leinass} that this additional postulate can be understood in terms of topological properties of the classical configuration
space of indistinguishable particles.

Let us denote by $M$ the one-particle classical configuration space (e.g., an $m$-dimensional manifold)
and by
\begin{eqnarray}
F_{n}(M)=\{(x_{1},\, x_{2},\ldots,\, x_{n})\,:x_{i}\in X,\, x_{i}\neq x_{j}\},
\end{eqnarray}
the space of $n$ distinct points in $M$. The $n$-particle
configuration space is defined as an orbit space
\begin{eqnarray}
C_{n}(M)=\nicefrac{F_{n}(M)}{S_{n}},
\end{eqnarray}
where $S_{n}$ is the permutation group of $n$ elements and the action
of $S_{n}$ on $F_{n}(M)$ is given by
\begin{eqnarray}
\sigma(x_{1}\,,\ldots\, x_{2})=(x_{\sigma(1)}\,,\ldots\, x_{\sigma(2)}),\,\,\,\,\forall\sigma\in S_{n}.
\end{eqnarray}
Any closed loop in $C_{n}(M)$ represents a process in which particles start at some particular configuration and end up in the same configuration modulo that they might have been exchanged. The space of all loops up to continuous deformations equipped with loop composition is the fundamental group $\pi_{1}(C_{n}(M))$ (see \cite{Hatcher} for more detailed
definition).




The abelianization of the fundamental group is the first homology group $H_{1}(C_{n}(M))$, and its structure plays an important role in the characterization of quantum statistics. In order to clarify this idea we will first consider the well-known problem of quantum statistics of many particles in $\mathbb{R}^{m}$, $m\geq2$. We will describe fully both the fundamental and homology groups of $C_{n}(\mathbb{R}^{m})$ for $m\geq2$, showing that for $m\geq3$, the only possible statistics are bosonic and fermionic, while for $m=2$ anyon statistics emerges. Next we pass to the main problem of this paper, namely $M=\Gamma$ is a quantum (metric) graph. We describe combinatorial structure of $C_{n}(\Gamma)$ and show how to compute $H_{1}(C_{n}(\Gamma))$ using discrete Morse theory.

\subsection{Quantum statistics for $C_{n}(\mathbb{R}^{m})$}

\paragraph{The case $M=\mathbb{R}^{m}$ and $m\geq3$.}

When $M=\mathbb{R}^{m}$ and $m\geq3$ the fundamental group $\pi_{1}(F_{n}(\mathbb{R}^{m}))$
is trivial, since there are enough degrees of freedom to avoid coincident configurations during the continuous contraction of any loop. Let us recall that we have a natural action of the permutation group $S_{n}$ on $F_{n}(\mathbb{R}^{m})$
which is free%
\footnote{The action of a group $G$ on $X$ is free iff the stabilizer of any $x\in X$
is the neutral element of $G$ .%
}. In such a situation the following theorem holds \cite{Hatcher}.

\begin{theo}If an action of a finite group $G$ on a space $Y$ is
free then $G$ is isomorphic to $\nicefrac{\pi_{1}(\nicefrac{Y}{G})}{p_{\ast}(\pi_{1}(Y))}$, where $p:Y\rightarrow\nicefrac{Y}{G}$ is the natural projection and
$p_{\ast}:\pi_{1}(Y)\rightarrow\pi_{1}(\nicefrac{Y}{G})$ is the induced map of fundamental groups.
\end{theo}

\noindent Notice that in particular if $\pi_{1}(Y)$ is trivial we
get $G=\pi_{1}(\nicefrac{Y}{G})$. In our setting $Y=F_{n}(\mathbb{R}^{m})$
and $G=S_{n}$. The triviality of $\pi_{1}(F_{n}(\mathbb{R}^{m}))$
implies that  the fundamental group of $C_{n}(\mathbb{R}^{m})$ is given
by
\begin{eqnarray}
\pi_{1}(\nicefrac{F_{n}(\mathbb{R}^{m})}{S_{n}})=\pi_{1}(C_{n}(\mathbb{R}^{m}))=S_{n}.
\end{eqnarray}
The homology group $H_{1}(C_{n}(\mathbb{R}^{m})\,,\,\mathbb{Z})$
is the abelianization of $\pi_{1}(C_{n}(\mathbb{R}^{m}))$.  Hence,
\begin{eqnarray}
H_{1}(C_{n}(\mathbb{R}^{m})\,,\,\mathbb{Z})=\mathbb{Z}_{2}.\label{eq:homr3}
\end{eqnarray}
Notice that $H_{1}(C_{n}(\mathbb{R}^{m})\,,\,\mathbb{Z})$ might also be
represented as $(\{1\,,\, e^{i\pi}\},\cdot)$. This result can explain
why we have only bosons and fermions in $\mathbb{R}^{m}$ when $m\geq3$ (see, e.g. \cite{Dowker85} for a detailed discussion).

\paragraph{The case $M=\mathbb{R}^{2}$.}

The case of $M=\mathbb{R}^{2}$ is different as $\pi_{1}(F_{n}(\mathbb{R}^{m}))$
is no longer trivial and it is hard to use Theorem 1 directly. In
fact it can be shown (see \cite{Fox}) that for $M=\mathbb{R}^{2}$
the fundamental group $\pi_{1}(C_{n}(\mathbb{R}^{2}))$ is Artin braid
group $\mathrm{Br}_{n}(\mathbb{R}^{2})$
\begin{eqnarray}
\mathrm{Br}_{n}(\mathbb{R}^{2})=\langle\sigma_{1},\sigma_{2},\ldots,\sigma_{n-1}\,|\,\sigma_{i}\sigma_{i+1}\sigma_{i}=\sigma_{i+1}\sigma_{i}\sigma_{i+1},\,\sigma_{i}\sigma_{j}=\sigma_{j}\sigma_{i}\rangle,
\end{eqnarray}
where in the first group of relations we take $1\leq i\leq n-2$,
and in the second, we take $|i-j|\geq2.$ Although this group
has a complicated structure, it is easy to see that its abelianization is
\begin{eqnarray}
H_{1}(C_{n}(\mathbb{R}^{2})\,,\,\mathbb{Z})=\mathbb{Z}.
\end{eqnarray}
This simple fact gives rise to a phenomena called anyon statistics \cite{leinass,wilczek},
i.e., particles in $\mathbb{R}^{2}$ are no longer fermions or bosons
but instead any phase $e^{i\phi}$ can be gained when they are exchanged \cite{Dowker85}.

\subsection{Quantum statistics for $C_{n}(\Gamma)$\label{sec:Quantum-statisctics-for-Cn}}

Let $\Gamma=(V\,,\, E)$ be a metric connected simple graph on $|V|$
vertices and $|E|$ edges. Similarly to the previous cases we
define
\begin{eqnarray}
F_{n}(\Gamma)=\{(x_{1},\, x_{2},\ldots,\, x_{n})\,:x_{i}\in\Gamma,\, x_{i}\neq x_{j}\},
\end{eqnarray}
and
\begin{eqnarray}
C_{n}(\Gamma)=F_{n}(\Gamma)/S_{n},
\end{eqnarray}
where $S_{n}$ is the permutation group of $n$ elements. Notice also
that the group $S_{n}$ acts freely on $F_{n}(\Gamma)$, which means that
$F_{n}(\Gamma)$ is the covering space of $C_{n}(\Gamma)$. In seems
a priori a difficult task to compute $H_{1}(C_{n}(\Gamma))$. Fortunately,
this problem can be reduced to the computation of the first homology group of some cell
complex, which we define now.

We begin with the notion of a cell complex \cite{Hatcher}. Let $B_{n}=\{x\in\mathbb{R}^{n}\,:\,\|x\|\leq1\}$
be the standard unit-ball. The boundary of $B_{n}$ is the unit-sphere
$S^{n-1}=\{x\in\mathbb{R}^{n}\,:\,\|x\|=1\}$. A cell complex $X$
is a nested sequence of topological spaces
\begin{eqnarray}
X^{0}\subseteq X^{1}\subseteq\dots\subseteq X^{n},
\end{eqnarray}
where the $X^{k}$'s are the so-called $k$ - skeletons defined as follows:
\begin{itemize}
\item The $0$ - skeleton $X^{0}$ is a discrete set of points.
\item For $\mathbb{N}\ni k>0$, the $k$ - skeleton $X^{k}$ is the result
of attaching $k$ - dimensional balls $B_{k}$ to $X^{k-1}$ by gluing
maps
\begin{eqnarray}
\sigma:S^{k-1}\rightarrow X^{k-1}.
\end{eqnarray}

\end{itemize}
By $k$-cell we understand the interior of the ball $B_{k}$ attached
to the $(k-1)$ - skeleton $X^{k-1}$. The $k$ - cell is regular if its
gluing map is an embedding (i.e., a homeomorphism onto its image).

Notice that every simple graph $\Gamma$ is a regular cell complex with
vertices as $0$-cells and edges as $1$-cells. If a graph contains
loops, these loops are irregular $1$ - cells (the two points that comprise the boundary of
$B_{1}$ are attached to a single vertex of  the $0$ - skeleton). The product
$\Gamma^{\times n}$ inherits a cell - complex structure; its cells
are cartesian products of cells of $\Gamma$. However, the spaces $F_{n}(\Gamma)$
and $C_{n}(\Gamma)$ are not cell complexes, as the points $\Delta=\{(x_{1},x_{2},\ldots,x_{n}):\exists_{i,j}\, x_{i}=x_{j}\}$
have been excised from them. Fortunately, there exists a cell complex which can be
obtained directly from $C_{n}(\Gamma)$ and which has the same homotopy
type.

Following \cite{Ghrist}
we define  the $n$-particle combinatorial configuration space as
\begin{eqnarray}
\mathcal{D}^n(\Gamma)=(\Gamma^{\times n}-\tilde{\Delta})/S_{n},
\end{eqnarray}
where $\tilde{\Delta}$ denotes all cells whose closure intersects
with $\Delta$. The space $\mathcal{D}^n(\Gamma)$ possesses a natural cell - complex
structure with vertices as $0$-cells, edges  as $1$-cells, $2$-cells corresponding to moving two particles
along two disjoint
edges in $\Gamma$, and $k$ - cells defined  analogously. The existence of a cell - complex structure happens to be very helpful for investigating
the homotopy structure of the underlying space. Namely, we have the following theorem:
\begin{theo}\label{Abrams_thm}\cite{Abrams,PS09} For any graph $\Gamma$
with at least $n$ vertices, the inclusion $\mathcal{D}^n(\Gamma)\hookrightarrow C_{n}(\Gamma)$
is a homotopy equivalence iff the following hold:
\begin{enumerate}
\item Each path between distinct vertices of valence not equal to two passes
through at least $n-1$ edges.
\item Each closed path in $\Gamma$ passes through at least $n+1$ edges.
\end{enumerate}
\end{theo}

\noindent For $n=2$  these conditions are automatically satisfied (provided $\Gamma$ is simple).  Intuitively, they can be understood as follows:
\begin{enumerate}
\item In order to have homotopy equivalence between $\mathcal{D}^n(\Gamma)$ and
$C_{n}(\Gamma)$, we need to be able to accommodate $n$ particles on
every edge of graph $\Gamma$.
\item For every cycle there is at least one free (not occupied) vertex which
enables the exchange of particles along this cycle.
\end{enumerate}
Using Theorem \ref{Abrams_thm}, the problem of finding  $H_{1}(C_{n}(\Gamma))$
is reduced to the problem of computing $H_{1}(\mathcal{D}^n(\Gamma))$. In the next sections we will discuss
how to determine $H_{1}(\mathcal{D}^n(\Gamma))$ using the discrete Morse
theory of Forman \cite{forman}. In order to clarify the idea behind theorem \ref{Abrams_thm} let us consider the following example.

\begin{ex}Let $\Gamma$ be a star graph on four vertices (see figure
1(a)). The two - particle configuration spaces $C_{2}(\Gamma)$ and
$\mathcal{D}^2(\Gamma)$ are shown in figures 1(b),(c). Notice that $C_{2}(\Gamma)$
consists of six $2$ - cells (three are interiors of triangles
and the other three are interiors of squares), eleven $1$ - cells and
six $0$ - cells. Vertices $(1,1)$, $(2,2)$, $(3,3)$ and $(4,4)$
do not belong to $C_{2}(\Gamma)$. Similarly dashed edges, i.e.~$(1,1)-(2,2)$,
$(2,2)-(4,4)$, $(2,2)-(3,3)$ do not belong to $C_{2}(\Gamma)$.
This is why $C_{2}(\Gamma)$ is not a cell complex - not every cell
has its boundary in $C_{2}(\Gamma)$. Notice that cells of $C_{2}(\Gamma)$ whose closures
intersect $\Delta$ (denoted by dashed lines and diamond points) do not
influence the homotopy type of $C_{2}(\Gamma)$ (see  figures 1(b),(c)). Hence, the space $\mathcal{D}^2(\Gamma)$ has
the same homotopy type as $C_{2}(\Gamma)$, but consists of six $1$
- cells and six $0$ - cells. $\mathcal{D}^2(\Gamma)$ is subspace of $C_{2}(\Gamma)$
denoted by dotted lines in figure 1(b).

\begin{figure}[H]
\includegraphics[scale=0.35]{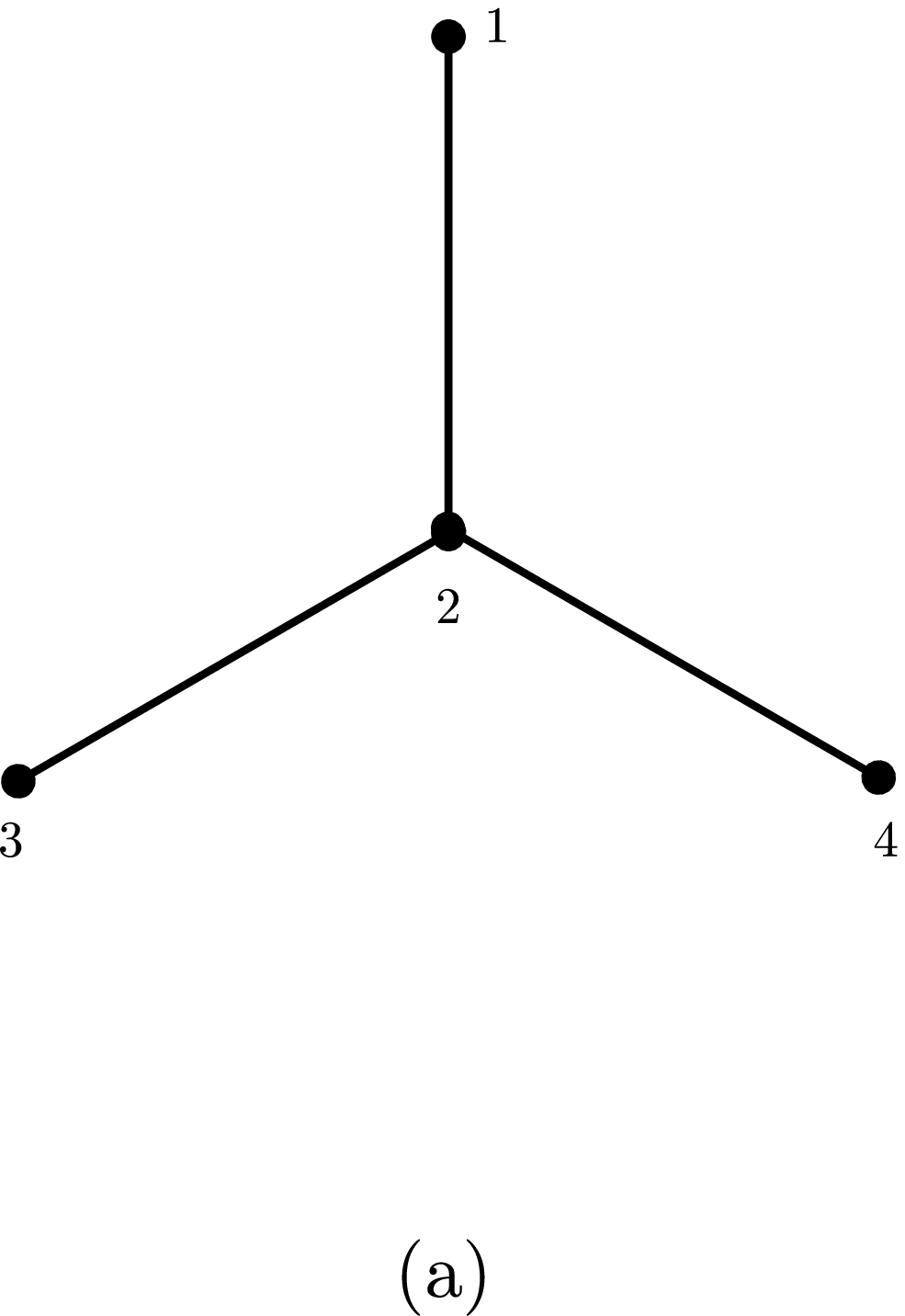}~~~\includegraphics[scale=0.35]{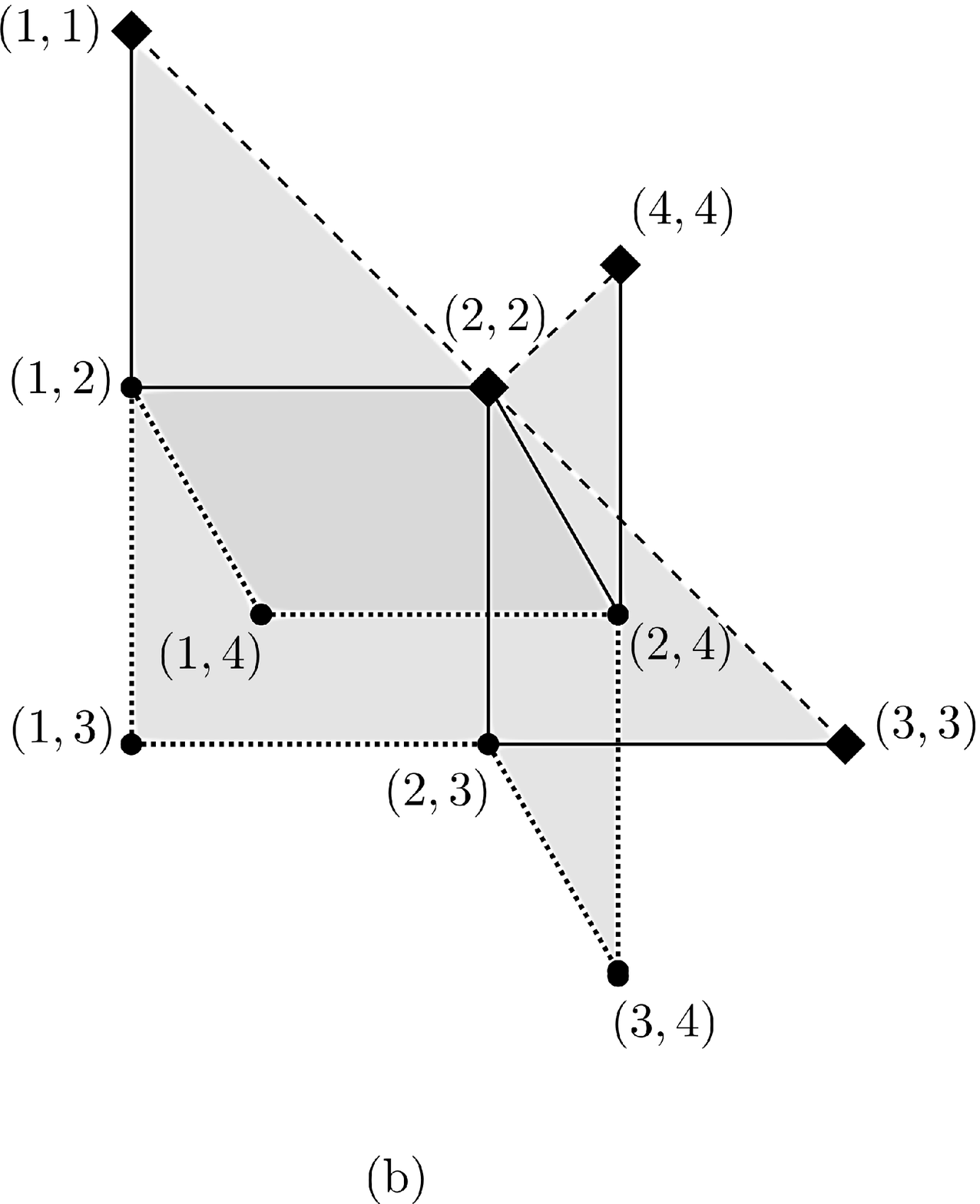}~~~\includegraphics[scale=0.35]{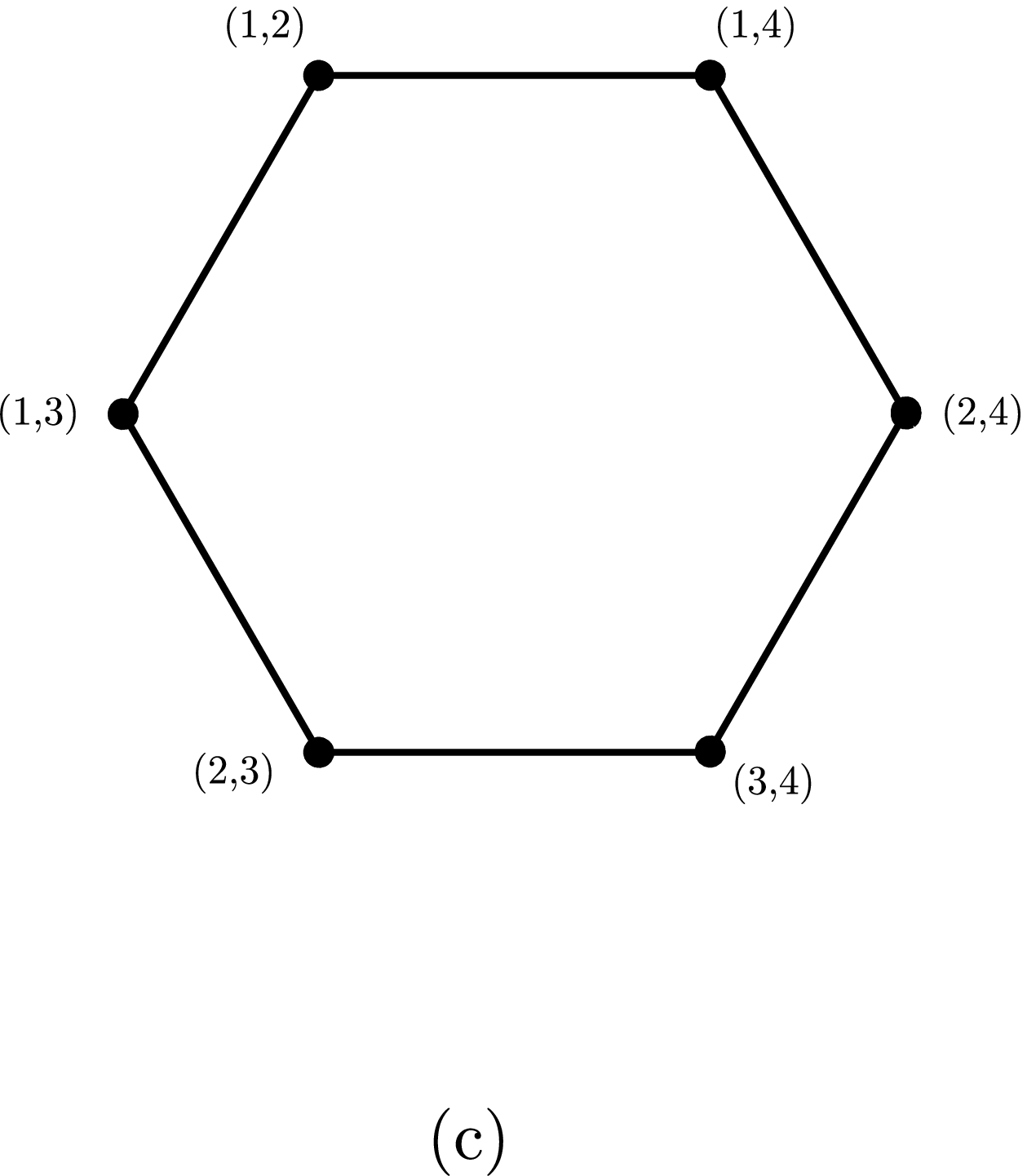}

\caption{(a) The star graph $\Gamma$, (b) the two-particle configuration space
$C_{2}(\Gamma)$, (c) the two-particle discrete configuration space
$\mathcal{D}^2(\Gamma)$.}
\end{figure}

\end{ex}

\section{Morse theory in the nutshell\label{sec:Morse-theory-in}}

In this section we briefly present both classical and discrete Morse
theories. We focus on the similarities between them and illustrate
the ideas by several simple examples.

\subsection{Classical Morse theory}

The concept of classical Morse theory is essentially very similar to its discrete counterpart. Since the former is better known we have found it beneficial to first discuss the classical version. A good reference is the monograph by
Milnor \cite{milnor}. Classical Morse theory is a useful tool
to describe topological properties of compact manifolds. Having such
a manifold $M$ we say that a smooth function $f:M\rightarrow\mathbb{R}$
is a Morse function if its Hessian matrix at every critical point
is nondegenerate, i.e.,
\begin{eqnarray}
df(x)=0\,\,\Rightarrow\,\,\mathrm{det}\left(\frac{\partial^{2}f}{\partial x_{i}\partial x_{j}}\right)(x)\neq0.
\end{eqnarray}
It can be shown that if $M$ is compact then $f$ has a finite number
of isolated critical points \cite{milnor}. The classical Morse theory
is based on the following two facts:
\begin{enumerate}
\item Let $M_{c}=\{x\in M\,:\, f(x)\leq c\}$ denote a sub level set of $f$.  Then $M_{c}$  is homotopy equivalent to $M_{c^{\prime}}$ if there is no critical value\footnote{A critical value of $f$  is the value of $f$ at one of its critical points.} between
the interval $(c,c\prime)$.
\item The change in topology when $M_{c}$ goes through a critical value
is determined by the index (i.e., the number of negative eigenvalues) of the Hessian
matrix at the associated critical point.
\end{enumerate}
The central point of classical Morse theory are the so-called Morse
inequalities, which relate the Betti numbers $\beta_{k}=\mathrm{dim}H_{k}(M)$, i.e. the dimensions of k-homology groups \cite{Hatcher}, to the numbers $m_{k}$ of critical points of index $k$, i.e.,
\begin{eqnarray}
\sum_{k}m_{k}t^{k}-\sum_{k}\beta_{k}t^{k}=(1+t)\sum_{k}q_{k}t^{k},\label{eq:Morse_ineq}
\end{eqnarray}
where $q_{k}\geq0$ and $t$ is an arbitrary real number. In particular
(\ref{eq:Morse_ineq}) implies that $\beta_{k}\leq m_{k}$. The function
$f$ is called a perfect Morse function iff $\beta_{k}=m_{k}$ for every
$k$. Since there is no general prescription it is typically hard to find a perfect Morse
function for a given manifold $M$. In fact a perfect Morse function may even not exist \cite{Ayala11}. However, even if $f$ is not perfect we can still encode the topological properties of $M$
in a quite small cell complex. Namely it follows from Morse theory that given a Morse function $f$,
one can show that $M$ is homotopic to a cell complex with $m_k$ $k$-cells,
and the gluing maps can be constructed in terms of the gradient paths of $f$.
We will not discuss this as it is far more complicated than in the discrete case.

\subsection{Discrete Morse function\label{sub:Discrete-Morse-function}}

In this section we discuss the concept of discrete Morse functions
for cell complexes as introduced by Forman \cite{forman}. Let $\alpha^{(p)}\in X$ denote a
$p$ - cell. A discrete Morse function on a regular cell complex $X$ is a function $f$
which assigns larger values to higher-dimensional cells with `local'
exceptions.
\begin{defi}
\label{Morse-fuction}A function $f\,:\, X\rightarrow\mathbb{R}$
is a discrete Morse function iff for every $\alpha^{(p)}\in X$ we
have
\begin{eqnarray}
\#\{\beta^{(p+1)}\supset\alpha\,:\, f(\beta)\leq f(\alpha)\}\leq1,\\
\#\{\beta^{(p-1)}\subset\alpha\,:\, f(\beta)\geq f(\alpha)\}\leq1.
\end{eqnarray}
\end{defi}
In other words, definition \ref{Morse-fuction} states that for any
$p$ - cell $\alpha^{(p)}$, there can be $\mathbf{at\,\, most}$
one $(p+1)$ - cell $\beta^{(p+1)}$ containing  $\alpha^{(p)}$ for which $f(\beta^{(p+1)})$ is less than or equal to  $f(\alpha^{(p)})$.  Similarly, there can be $\mathbf{at\,\, most}$ one $(p-1)$ - cell $\beta^{(p-1)}$ contained in  $\alpha^{(p)}$ for which $f(\beta^{(p-1)})$ is greater than or equal to  $f(\alpha^{(p)})$. Examples of a Morse function and a non-Morse function are shown in figure \ref{figure3}. The most important part of discrete Morse theory is the definition of a critical cell:
\begin{defi}
\label{criticalcell}A cell $\alpha^{(p)}$ is critical iff
\begin{eqnarray}
\#\{\beta^{(p+1)}\supset\alpha\,:\, f(\beta)\leq f(\alpha)\}=0,\,\,\mathrm{and}\\
\#\{\beta^{(p-1)}\subset\alpha\,:\, f(\beta)\geq f(\alpha)\}=0.
\end{eqnarray}
\end{defi}
That is,  $\alpha$ is critical if  $f(\alpha)$ is greater than the value of $f$ on all of the faces of $\alpha$, and $f(\alpha)$ is greater than the value of $f$ on all cells containing $\alpha$ as a face. From definitions
\ref{Morse-fuction} and \ref{criticalcell}, we get that a cell $\alpha$ is noncritical iff either
\begin{enumerate}
\item $\exists \ {\rm unique}\ \tau^{(p+1)}\supset\alpha\,\,\,\,\, \ {\rm with}\  f(\tau)\leq f(\alpha),$
or
\item $\exists \ {\rm unique}\  \beta^{(p-1)}\subset\alpha\,\,\,\,\,\ {\rm with}\  f(\beta)\geq f(\alpha).$
\end{enumerate}
It is quite important to understand that these two conditions cannot
be simultaneously fulfilled, as we now explain. Let us assume
on the contrary that both conditions (i) and (ii) hold. We have the following
sequence of cells:
\begin{eqnarray}
\tau^{(p+1)}\supset\alpha^{(p)}\supset\beta^{(p-1)}.
\end{eqnarray}
Since $\alpha^{(p)}$ is regular there is necessarily an $\tilde{\alpha}^{(p)}$
such that $\tau^{(p+1)}\supset\tilde{\alpha}^{(p)}\supset\beta^{(p-1)}$
(see figures 2(a),(b) for an intuitive explanation). Since $f(\tau)\leq f(\alpha)$,
by definition \ref{Morse-fuction} we have
\begin{eqnarray}
f(\tilde{\alpha})<f(\tau).
\end{eqnarray}
We also know that $f(\beta)\geq f(\alpha)$ which, once again by definition
\ref{Morse-fuction}, implies $f(\beta)<f(\tilde{\alpha})$. Summing
up we get
\begin{eqnarray}
f(\alpha)\leq f(\beta)< f(\tilde{\alpha})<f(\tau)\leq f(\alpha),
\end{eqnarray}
which is a contradiction.
\begin{figure}[h]
~~~~~~~~~~~~~~~~~\includegraphics[scale=0.5]{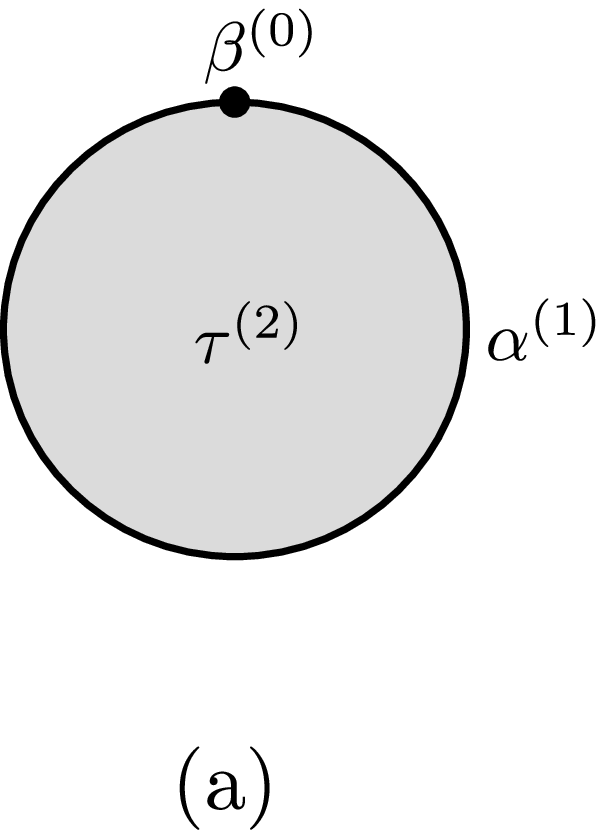}~~~~~~~~~~~~~~~~~~~~~~\includegraphics[scale=0.5]{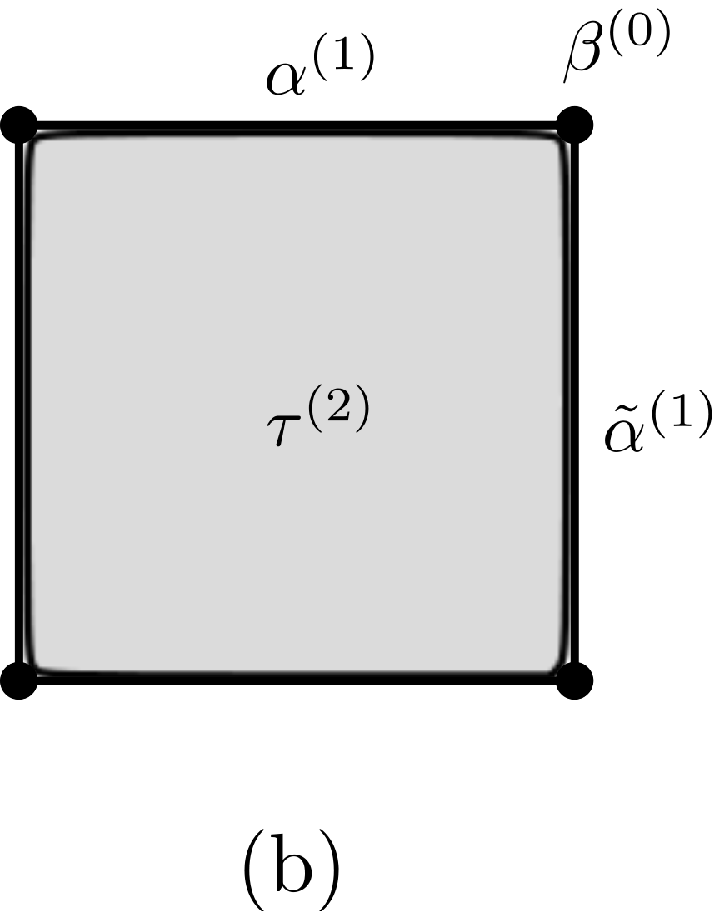}

\caption{Examples of (a) an irregular cell complex.  $\alpha^{(1)}$ is an irregular
1 - cell and $\beta^{(0)}$ is an irregular face of $\alpha^{(1)}$. (b) A
regular cell complex with $\tau^{(2)}\supset\alpha^{(1)}\supset\beta^{(0)}.$}
\end{figure}

\noindent Following the path of classical Morse theory we define
next the level sub-complex $K(c)$ by
\begin{eqnarray}
K(c)=\cup_{f(\alpha)\leq c}\cup_{\beta\subseteq\alpha}\beta.\label{eq:levelsubcomplex}
\end{eqnarray}
That is,  $K(c)$ is the sub-complex containing all cells on which $f$
is less or equal to $c$, $\mathbf{{together\,\, with\,\, their\,\, faces}}$%
\footnote{Notice that the value of $f$ on some of these faces might be bigger than
$c$.%
}.
Notice that by definition (\ref{Morse-fuction}) a Morse function does not have to be a bijection.  However, we have the following  \cite{forman}:
\begin{lem}\label{lem: Morse 1-1}
For any Morse function $f_1$, there exist another Morse function $f_2$ which
is 1-1 and which has the same critical cells as  $f_1$.
\end{lem}
The process of  attaching cells is accompanied by two important lemmas
which describe the change in homotopy type of level sub-complexes
when critical or noncritical cells are attached. Since, from lemma~\ref{lem: Morse 1-1}, we can assume that a given Morse function is 1-1, we can always choose the intervals $[a,b]$
below so that $f^{-1}([a,b])$ contains exactly one cell.
%
%
\begin{lem}
\label{lem:3} \cite{forman} If there are no critical cells $\alpha$
with $f(\alpha)\in[a,b]$, then $K(b)$ is homotopy equivalent to $K(a)$.
\end{lem}

\begin{lem}
\label{lem:4}\cite{forman} If there is a single critical cell $\alpha^{(p)}$
with $f(\alpha)\in[a,b]$, then $K(b)$ is homotopy equivalent to
\begin{eqnarray}
K(b)=K(a)\cup\alpha
\end{eqnarray}
and $\partial\alpha\subset K(a)$.
\end{lem}
The above two lemmas lead to the following conclusion:
\begin{theo}
\label{thm:1}\cite{forman} Let $X$ be a cell complex and $f\,:\, X\rightarrow\mathbb{R}$
be a Morse function. Then $X$ is homotopy equivalent to a cell complex
with exactly one cell of dimension $p$ for each critical cell $\alpha^{(p)}$

\begin{figure}[h]\label{figure3}
~~~~~~~~~~~~~~~~~\includegraphics[scale=0.5]{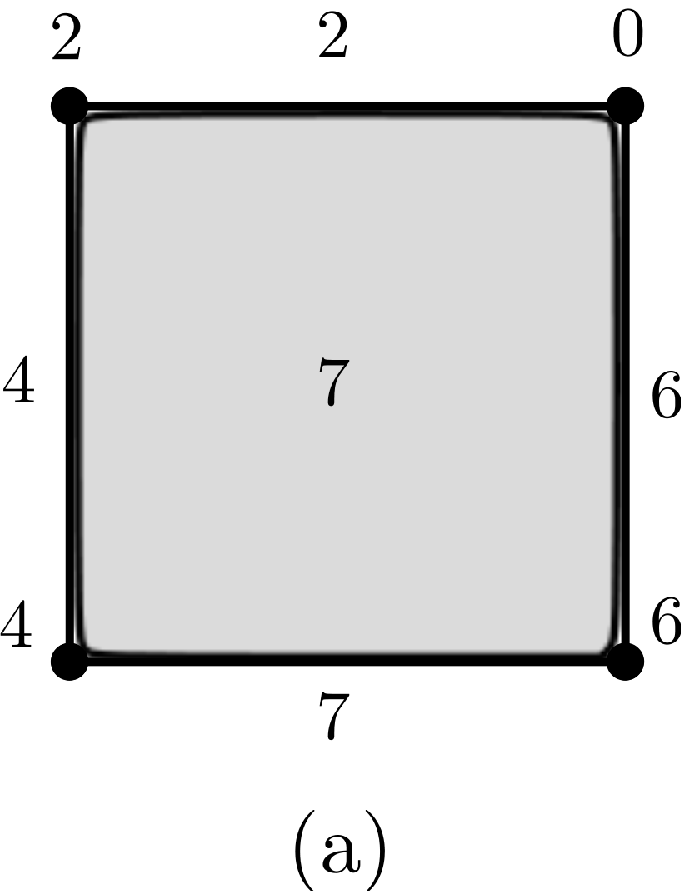}~~~~~~~~~~~~~~~~~~~~\includegraphics[scale=0.5]{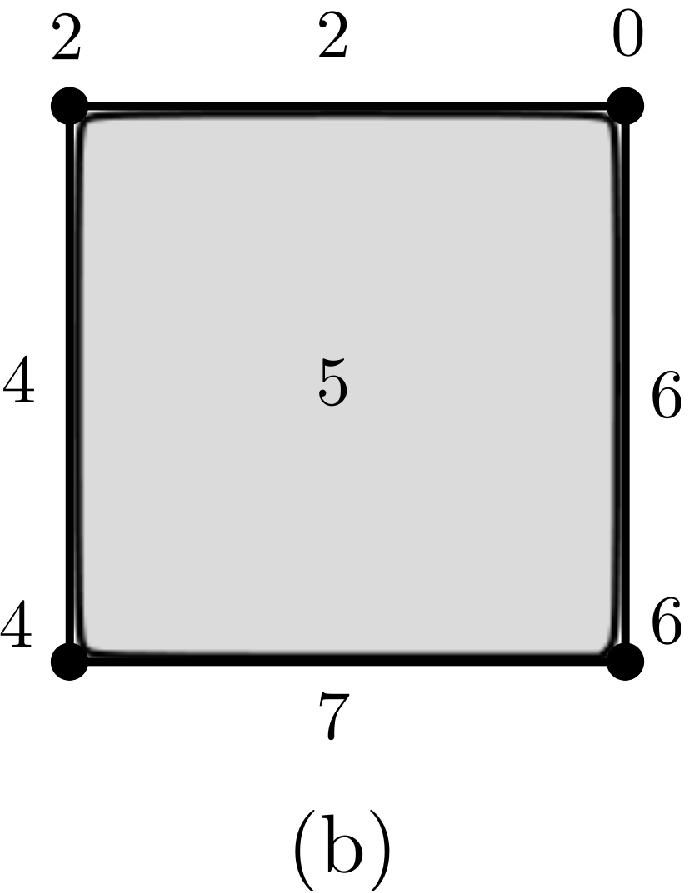}

\caption{Examples of (a) a Morse function, and (b) a non-Morse function, since the 2-cell has value $5$ and there are two $1$-cells in its boundary
with higher values assigned ($6$, $7$).}
\end{figure}

\end{theo}

\subsection{Discrete Morse vector field}
From theorem \ref{thm:1} it follows that a given cell complex is homotopy equivalent to a  cell complex containing only its critical cells, the so-called Morse complex. The construction of the Morse complex, in particular its boundary map (as well as the proof of theorem~\ref{thm:1}), depends crucially on the concept of a discrete vector field, which  we define next.
We know from definition \ref{Morse-fuction} that the noncritical
cells can be paired. If a $p$-cell is noncritical, then it is paired
with either the unique noncritical $(p+1)$-cell on which $f$ takes
an equal or smaller value, or the unique noncritical $(p-1)$-cell on
which $f$ takes an equal or larger value. In order to indicate this
pairing we draw an arrow from the $(p-1)$-cell to the $p$-cell in the first
case or from the $p$-cell to the $(p+1)$-cell in the second case (see figure
4). Repeating this for all cells we get the so-called discrete gradient
vector field of the Morse function. It also follows from section \ref{sub:Discrete-Morse-function}
that for every cell $\alpha$ exactly one of the following is true:
\begin{enumerate}
\item $\alpha$ is the tail of one arrow,
\item $\alpha$ is the head of one arrow,
\item $\alpha$ is neither the tail nor the head of an arrow.
\end{enumerate}
Of course $\alpha$ is critical iff it is neither the tail nor the
head of an arrow. Assume now that we are given a collection of arrows
on some cell complex satisfying the above three conditions. The question
we would like to address is whether it is a gradient vector field
of some Morse function. In order to answer this question we need to
be more precise. We define
\begin{defi}
A discrete vector field $V$ on a cell complex $X$ is a collection
of pairs $\{\alpha^{(p)}\subset\beta^{(p+1)}\}$ of cells such that
each cell is in at most one pair of $V$.
\end{defi}
Having a vector field it is natural to consider its `integral lines'.
We define the $V$ - path as a sequence of cells
\begin{eqnarray}
\alpha_{0}^{(p)},\,\beta_{0}^{(p+1)},\,\alpha_{1}^{(p)},\,\beta_{1}^{(p+1)},\,\ldots,\alpha_{k}^{(p)},\,\beta_{k}^{(p+1)}\label{eq:Vpath}
\end{eqnarray}
such that$\{\alpha_{i}^{(p)}\subset\beta_{i}^{(p+1)}\}\in V$ and
$\beta_{i}^{(p+1)}\supset\alpha_{i+1}^{(p)}$. Assume now that $V$
is a gradient vector field of a discrete Morse function $f$ and consider
a $V$ - path (\ref{eq:Vpath}). Then of course we have
\begin{eqnarray}
f(\alpha_{0}^{(p)})\geq f(\beta_{0}^{(p+1)})>f(\alpha_{1}^{(p)})\geq f(\beta_{1}^{(p+1)})>\ldots>f(\alpha_{1}^{(p)})\geq f(\beta_{k}^{(p+1)}).
\end{eqnarray}
This implies that if $V$ is a gradient vector field of
the Morse function then $f$ decreases along any $V$-path which in
particular means that there are no closed $V$-paths. It happens that
the converse is also true, namely a discrete vector field $V$ is a gradient
vector field of some Morse function iff there are no closed $V$ -
paths \cite{forman}.

\begin{figure}[h]
~~~~~~~~~~~~~~~~~\includegraphics[scale=0.5]{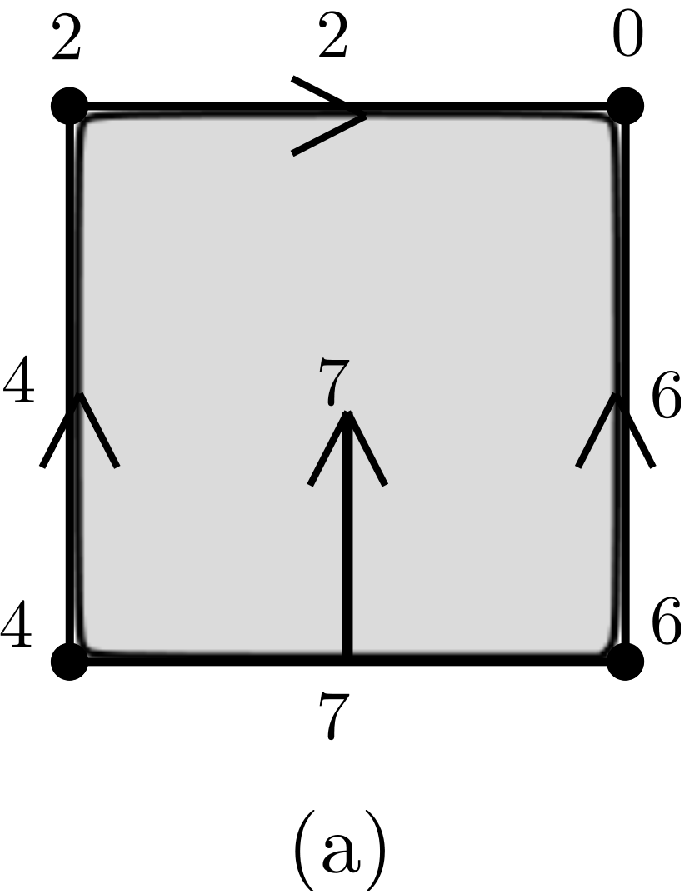}~~~~~~~~~~~~~~~~~~~~~~~~\includegraphics[scale=0.5]{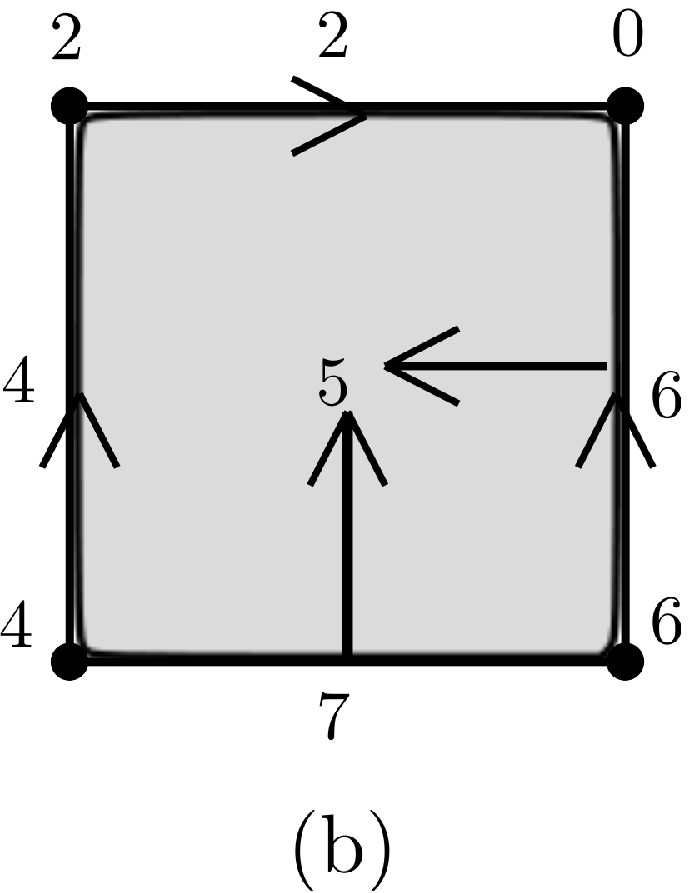}

\caption{Examples of (a) a correct and (b) an incorrect discrete gradient vector fields; the 2-cell is the head of two arrows and the 1-cell is the head and tail of one arrow.}
\end{figure}

\subsection{The Morse complex}

Up to now we have learned how to reduce the number of cells of the original
cell complex to the critical ones. However, it is still not clear
how these cells are `glued' together, i.e. what is the boundary map
between the critical cells? The following result relates the concept of critical cells with discrete gradient vector fields \cite{forman}.
\begin{theo}
Assume that orientation has been chosen for each cell in the cell complex
$X$. Then for any critical $(p+1)$-cell $\beta$ we have
\begin{eqnarray}
\tilde{\partial}\beta=\sum_{critical\,\alpha^{(p)}}c_{\beta,\alpha}\alpha,\label{eq:boundary}
\end{eqnarray}
where $\tilde{\ensuremath{\partial}}$ is the boundary map
in the cell complex consisting of the critical cells, whose existence
is guaranteed by theorem \ref{thm:1}, and
\begin{eqnarray}
c_{\beta,\alpha}=\sum_{\gamma\in P(\beta,\alpha)}m(\gamma),
\end{eqnarray}
where $P(\beta,\alpha)$ is the set of all $V$ - paths from
the boundary of $\beta$ to cells whose boundary contains $\alpha$ and
$m(\gamma)=\pm1$, depending on whether the orientation induced from
$\beta$ to $\alpha$ through $\gamma$ agrees with the one chosen for
$\alpha$.
\end{theo}
The collection of critical cells together with the boundary map $\tilde{\partial}$ is called the Morse complex of the function $f$ and we will denote it by $M(f)$. Examples of the computation of boundary maps for Morse complexes will be given in section \ref{sec:Main-example}.

\section{A perfect Morse function on $\Gamma$ and its discrete vector field.\label{sec:One-particle-graph}}

In this section we present a construction of a perfect discrete Morse function
on a $1$ - particle graph. It is defined analogously as in the classical case, i.e. the number of critical cells in each dimension is equal to the corresponding dimension of the homology group. The existence of such a function will
be used in section 5 to construct a `good' but not necessarily perfect
Morse function on a $2$-particle graph.

Let $\Gamma=(V\,,\, E)$ be a graph with $v=|V|$ vertices and $e=|E|$
edges. In the following we assume that $\Gamma$ is connected and
simple. Let $T$ be the spanning tree of $\Gamma$, i.e. $T$ is a connected
spanning subgraph of $\Gamma$ such that $V(T)=V(\Gamma)$ and for
any pair of vertices $v_{i}\neq v_{j}$ there is exactly one path
in $T$ joining $v_{i}$ with $v_{j}$. We naturally have $|E(\Gamma)|-|E(T)|\geq0$.
The Euler characteristic of $\Gamma$ treated as a cell
complex is given by
\begin{eqnarray}
\chi(\Gamma)=v-e=\mathrm{dim}H_{0}(\Gamma)-\mathrm{dim}H_{1}(\Gamma)=b_{0}-b_{1}.
\end{eqnarray}
Since $\Gamma$ is connected, $H_{0}(\Gamma)=\mathbb{Z}$. Hence we get
\begin{eqnarray}
b_{0}=1,\\
b_{1}=e-v+1.
\end{eqnarray}
On the other hand it is well known that $b_{1}=|E(\Gamma)|-|E(T)|$. Summing up from
the topological point of view $\Gamma$ is homotopy equivalent to
a wedge sum of $b_{1}$ circles. Our goal is to construct a perfect
Morse function $f_{1}$ on $\Gamma$, i.e. the one with exactly $b_{1}$
critical $1$ - cells and one critical $0$ - cell. To this end we choose a vertex $v_{1}$ of valency one in
$T$ (it always exists) and travel through the tree anticlockwise
from it labeling vertices by $v_{k}$. The value of $f$ on the vertex
$v_{k}$ is $f_{1}(k)=2k-2$ and the value of $f_{1}$ on the edge
$(i,j)\in T$ is $f_{1}((i,j))=\mathrm{max}\left(f_{1}(i),\, f_{1}(j)\right)$.
The last step is to define $f_{1}$ on the deleted edges $(i,j)\in E(\Gamma)\setminus E(T)$.
We choose $f_{1}((i,j))=\mathrm{max}(f_{1}(i),\, f_{1}(j))+2$, where
$v_{i},v_{j}$ are the boundary vertices of $(i,j)$. This way we obtain
that all vertices besides $v_{1}$ and all edges of $T$ are not critical
cells of $f_{1}$. The critical $1$ - cells are exactly the deleted edges.
The following example clarifies this idea (see figure 5).
\begin{ex}
Consider the graph $\Gamma$ shown in figure 5(a). Its spanning tree
is denoted by solid lines and the deleted edges by dashed lines. For each vertex and
edge the corresponding value of a perfect discrete Morse function
$f_1$ is explicitly written. Notice that according to definition
\ref{criticalcell} we have exactly one critical $0$ - cell (denoted
by a square) and four critical $1$ - cells which are deleted
edges. The discrete vector field for $f_1$ is represented by arrows.
The contraction of $\Gamma$ along this field yields the contraction of $T$ to a single
point and hence the Morse complex $M(f_1)$ is the wedge sum of four circles (see figure 5(b))
\end{ex}
\begin{figure}[H]
~~~~~~~~~~~\includegraphics[scale=0.43]{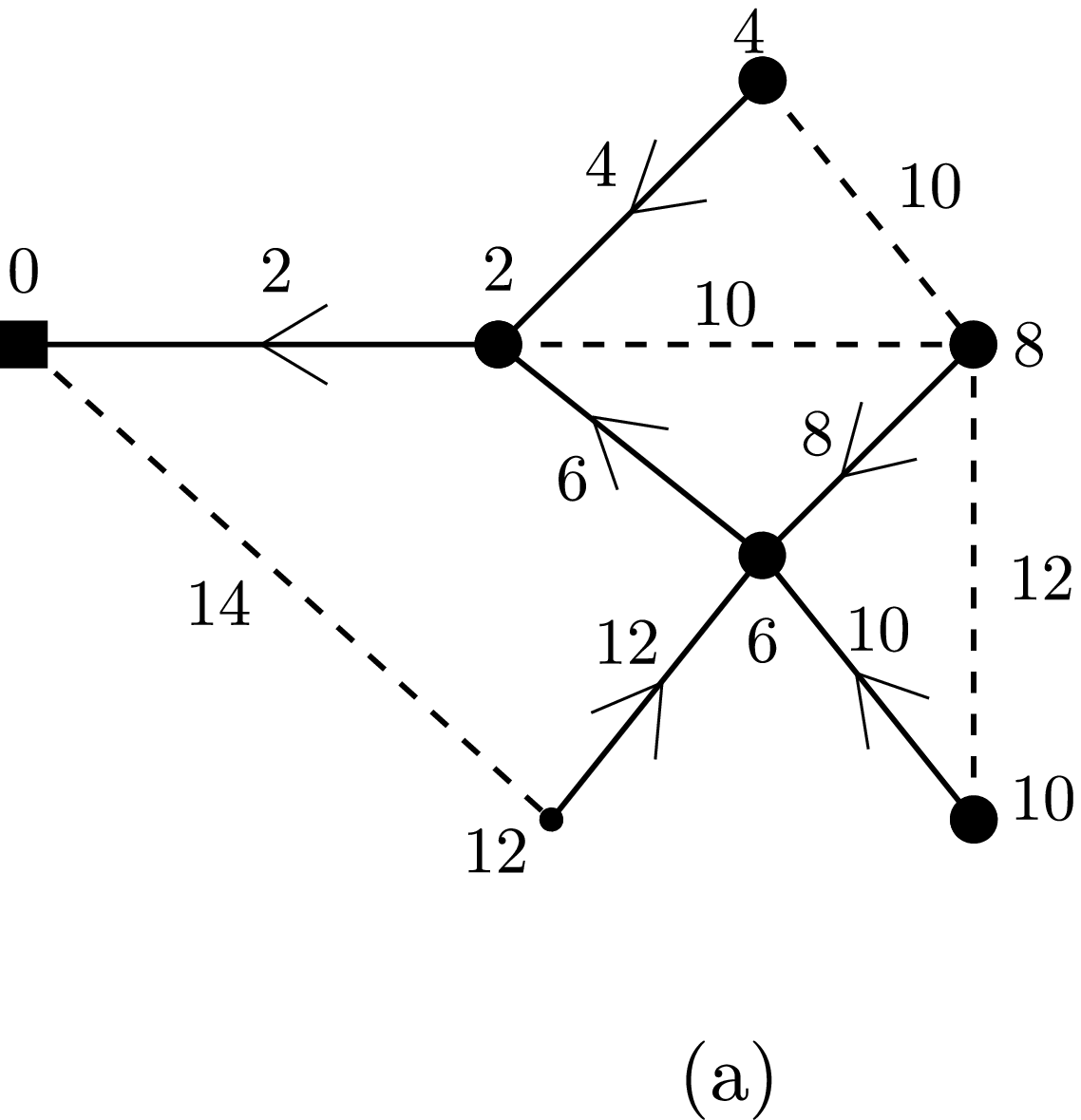}~~~~~~~~~~~~~~~~~\includegraphics[scale=0.5]{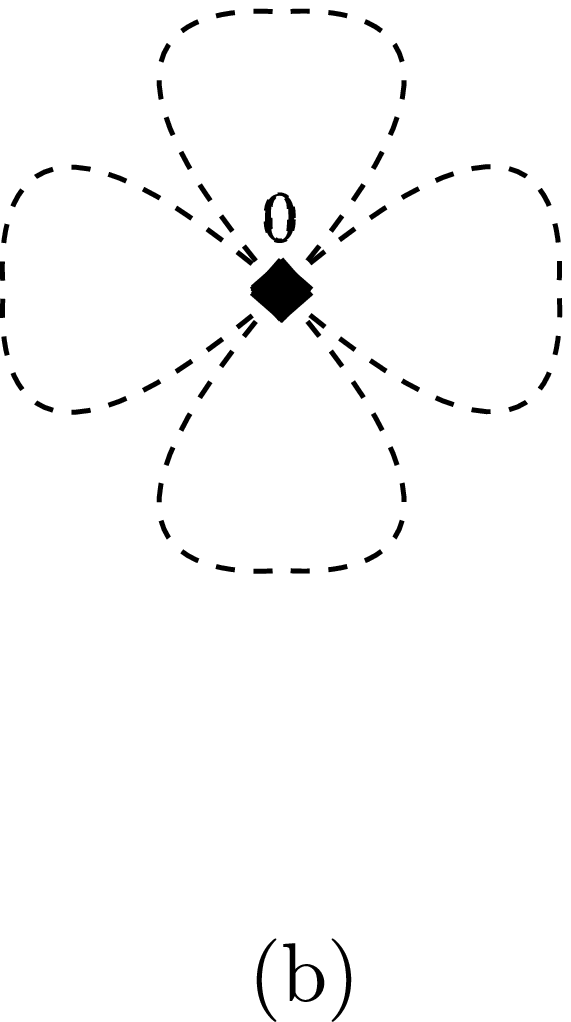}

\caption{(a) The perfect discrete Morse function $f_{1}$ on the graph $\Gamma$
and its discrete gradient vector field. (b) The Morse complex $M(f_{1})$.}
\end{figure}

\section{The main examples \label{sec:Main-example}}

In this section we present a method of construction of a `good' Morse
function on the two particle configuration space $\mathcal{D}^2(\Gamma_{i})$
for two different graphs $\Gamma_{i}$ shown in figures 6(a) and 8(a).
We also demonstrate how to use the tools described in section \ref{sec:Morse-theory-in}
in order to derive a Morse complex and compute the first homology group.
We begin with a graph $\Gamma_{1}$ which we will refer to as lasso
(see figure 6(a)). The spanning tree of $\Gamma_{1}$ is denoted in
black in figure 6(a). In figure 6(b) we see an example of the perfect
Morse function $f_{1}$ on $\Gamma_{1}$ together with its gradient
vector field. They were constructed according to the procedure explained
in section \ref{sec:One-particle-graph}. The Morse complex of $\Gamma_{1}$
consists of one $0$-cell (the vertex $1$) and one $1$-cell (the
edge $(3,4)$).

\begin{figure}[h]
~~~~~~~~~~~~~\includegraphics[scale=0.4]{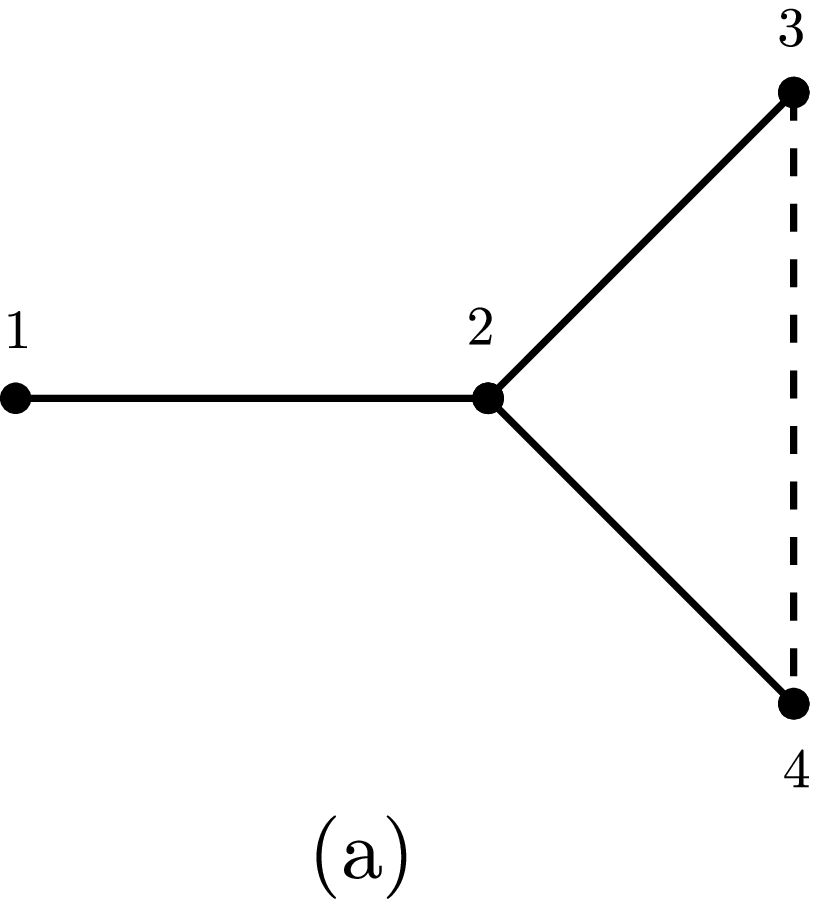}~~~~~~~~~~~~~~~~~~~~~~~~~~~~~~~~~~~~~\includegraphics[scale=0.4]{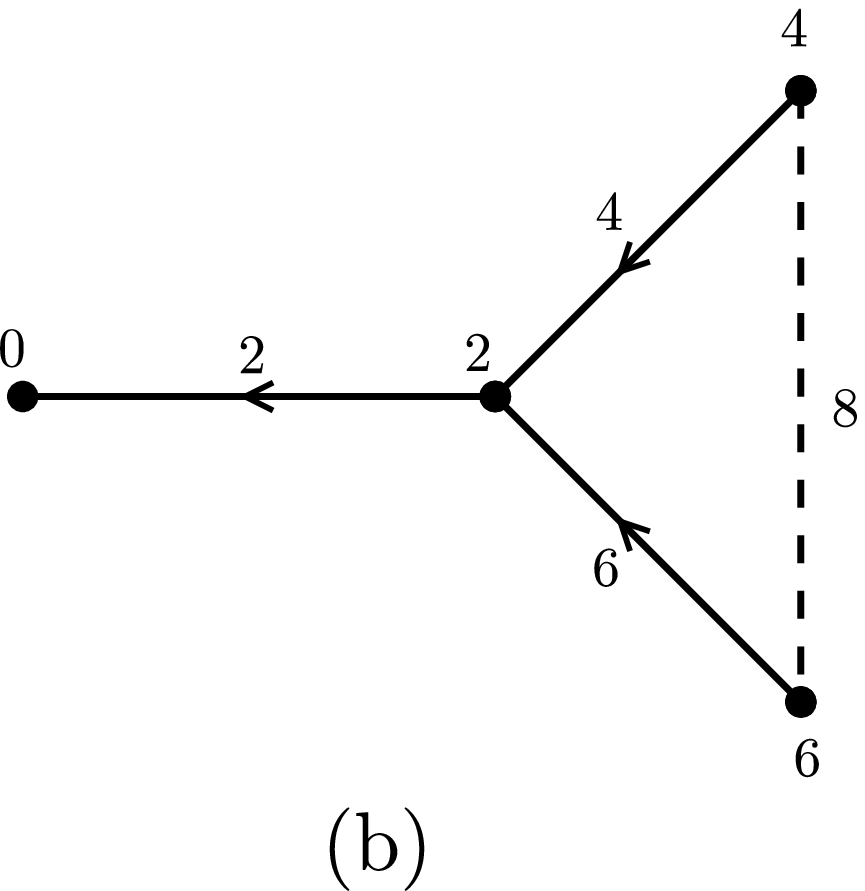}

\caption{(a) One particle on lasso, (b) The perfect discrete Morse function
$f_{1}$}
\end{figure}

\noindent The two particle configuration space $\mathcal{D}^2(\Gamma_{1})$
is shown in figure 7(a). Notice that $\mathcal{D}^2(\Gamma_{1})$ consists
of one $2$ - cell $(3,4)\times(1,2)$\footnote{This notation should be understood as the Cartesian product of edges $(3,4)$ and $(1,2)$, hence a square.}, six $0$ - cells and eight
$1$ - cells. In order to define the Morse function $f_{2}$ on $\mathcal{D}^2(\Gamma_{1})$
we need to specify its value for each of these cells. We begin with
a trial function $\tilde{f}_{2}$ which is completely determined once
we know the perfect Morse function on $\Gamma_{1}$. To this end
we treat $f_{1}$ as a kind of `potential energy' of one particle.
The function $\tilde{f}_{2}$ is simply the sum of the energies of both
particles, i.e. the value of $\tilde{f}_{2}$ on a cell corresponding
to a particular position of two particles on $\Gamma_{1}$ is the
sum of the values of $f_{1}$ corresponding to this position. To
be more precise we have for
\begin{eqnarray}
\mathrm{0-cells:\,\,\,\,\,\,\,\,\,\,\,\,\,}\,\,\,\,\,\,\,\,\,\,\,\,\tilde{f}_{2}(i\times j) & = & f_{1}(i)+f_{1}(j),\nonumber \\
\mathrm{1-cells:}\,\,\,\,\,\,\,\,\,\,\,\,\tilde{f}_{2}\left(i\times(j,k)\right) & = & f_{1}(i)+f_{1}\left((j,k)\right),\nonumber \\
\mathrm{2-cells:}\,\,\,\tilde{f}_{2}\left((i,j)\times(k,l)\right) & = & f_{1}\left((i,j)\right)+f_{1}\left((k,l)\right).\label{eq:rules}
\end{eqnarray}
In figure 7(b) we can see $\mathcal{D}^2(\Gamma_{1})$ together with $\tilde{f}_{2}$.
Observe that $\tilde{f}_{2}$ is not a Morse function since the value
of $\tilde{f}_{2}\left((3,4)\right)$ is the same as the value of
$\tilde{f}_{2}$ on edges $4\times(2,3)$ and $3\times(2,4)$ which
are adjacent to the vertex $(3,4)$. The rule that $0$ - cell can
be the face of at most one $1$ - cell with smaller or equal value of $\tilde{f}_{2}$
is violated. In order to have Morse function $f_{2}$ on $\mathcal{D}^2(\Gamma_{1})$
we introduce one modification, namely
\begin{eqnarray}
f_{2}\left(3\times(2,4)\right)=\tilde{f}_{2}\left(3\times(2,4)\right)+1,\label{eq:mod2}
\end{eqnarray}
and $f_{2}$ is $\tilde{f_{2}}$ on the other cells.
\begin{figure}[h]
~~~~\includegraphics[scale=0.4]{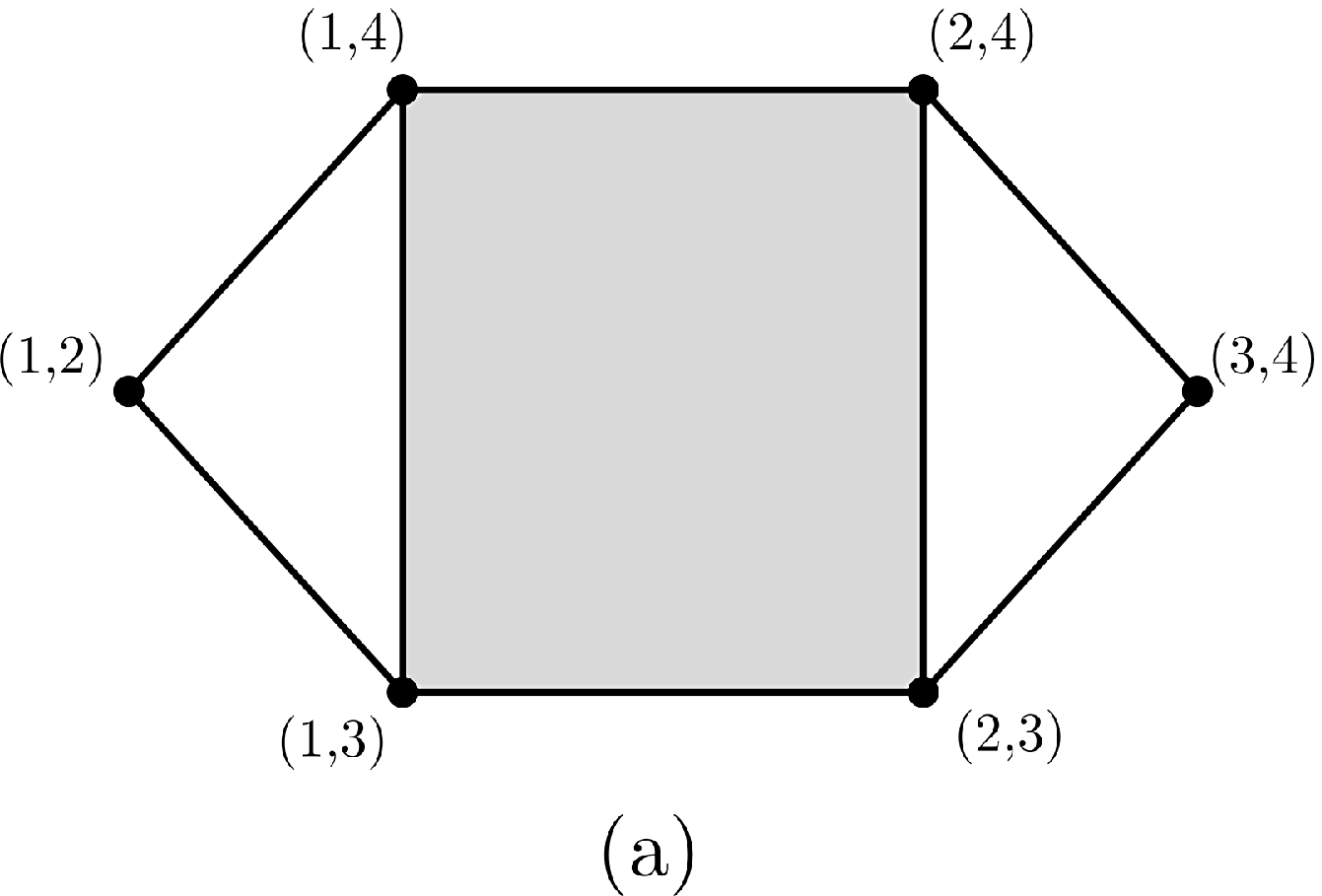}~~~~~~~~~\includegraphics[scale=0.4]{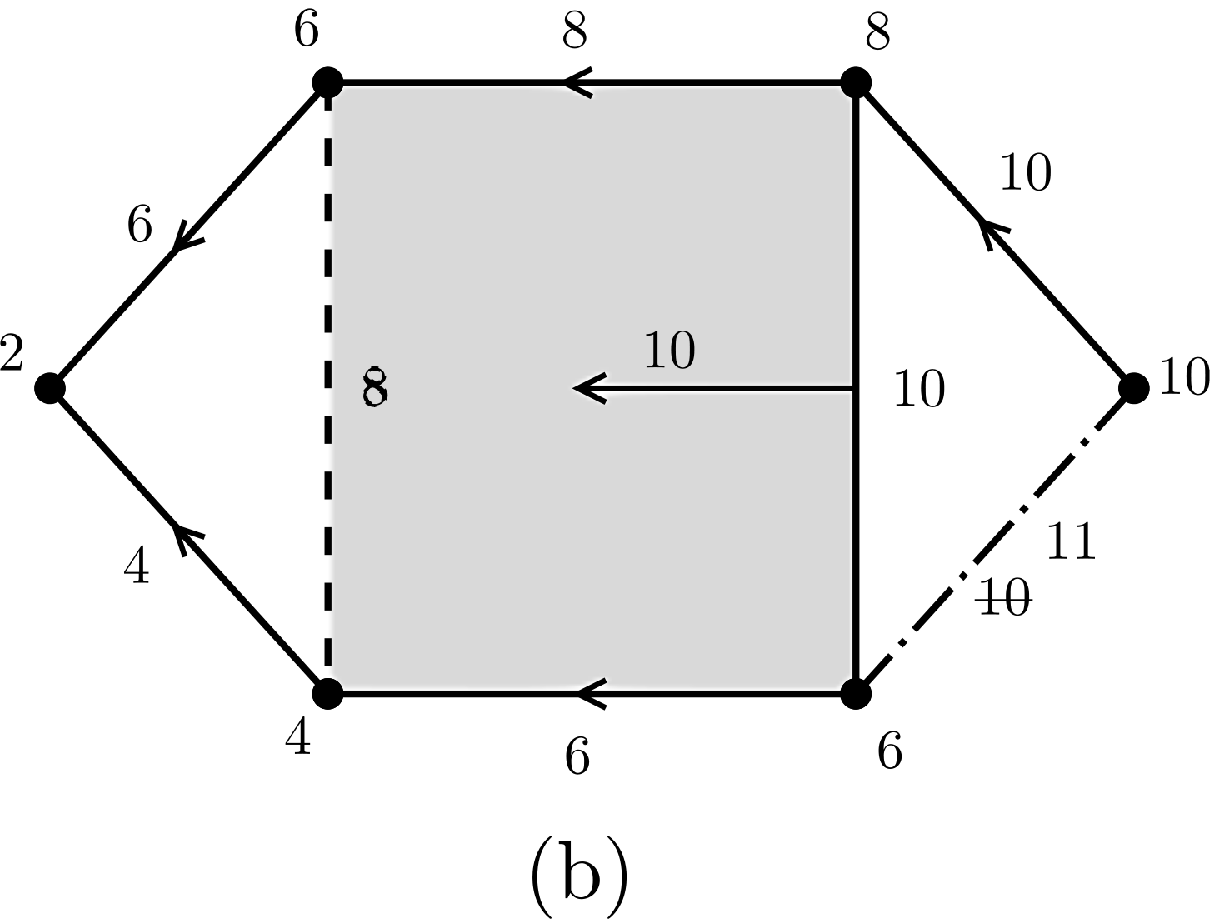}~~~~~~~~

\medskip{}

~~~~~~~~\includegraphics[scale=0.35]{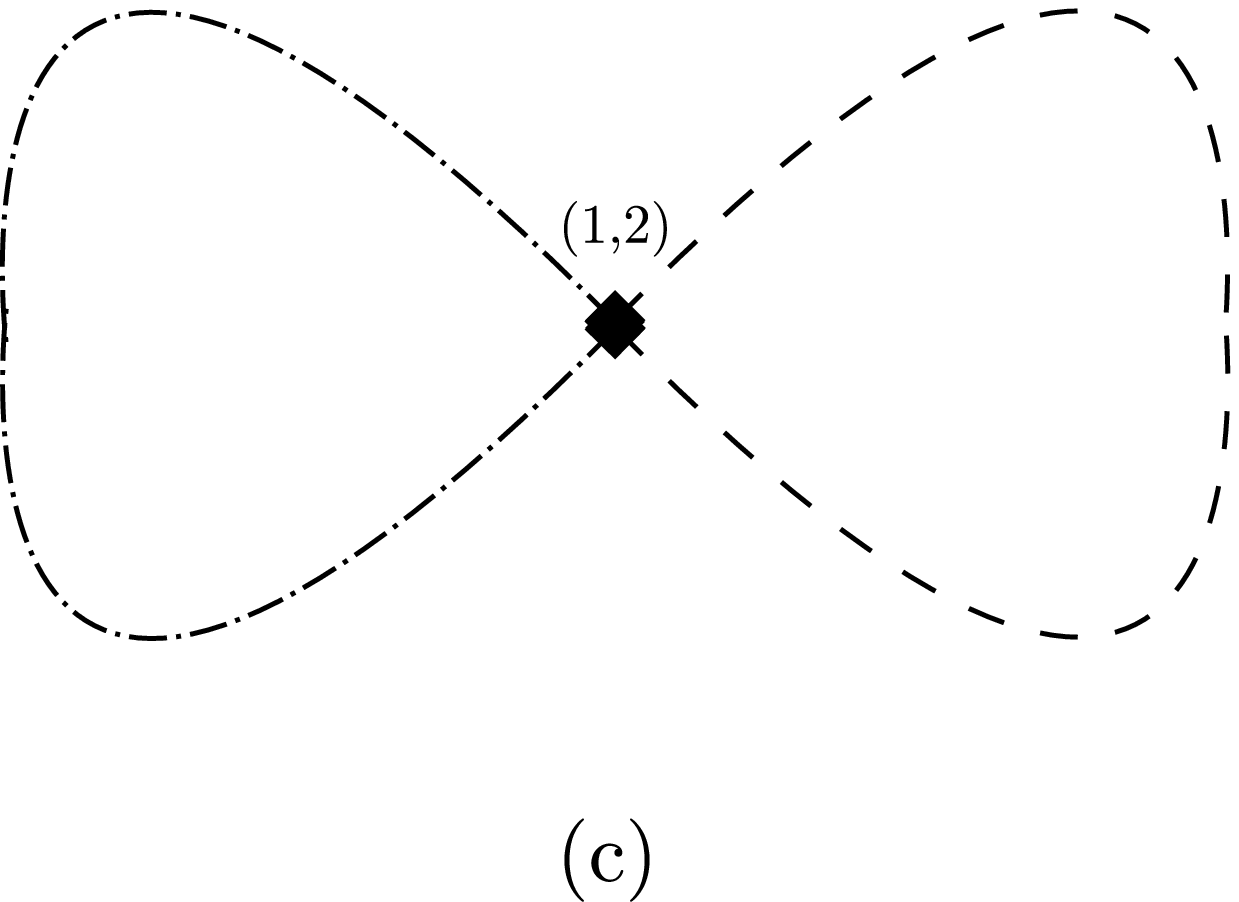}~~~~~~~~~~~\includegraphics[scale=0.4]{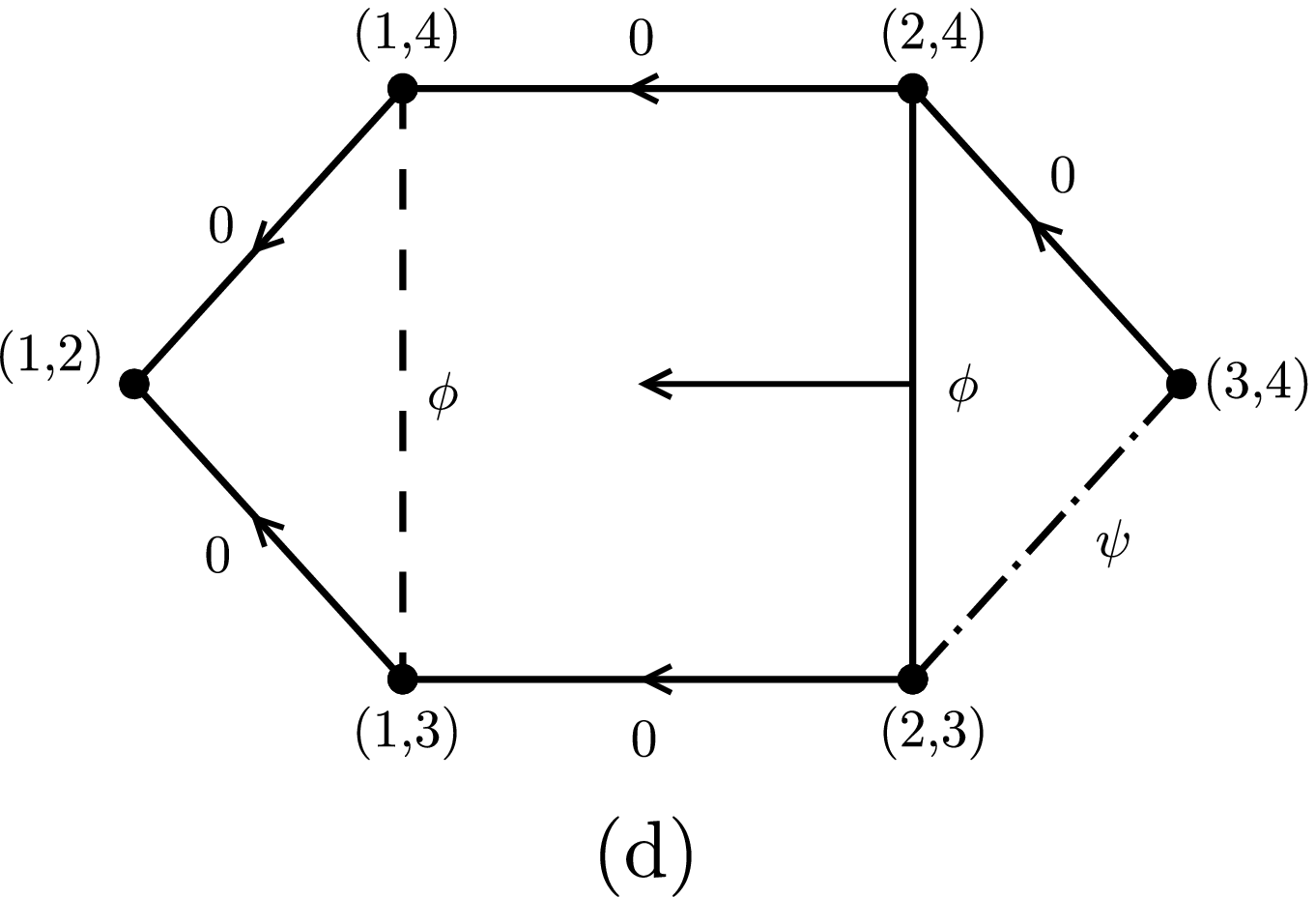}

\caption{(a) The two particles on lasso, $\mathcal{D}^2(\Gamma_{1})$, (b) the discrete Morse
function and its gradient vector field (c) the Morse complex (d) the topological gauge potential $\Omega$}
\end{figure}

\noindent Notice that the choice we made is not unique. We could have changed $\tilde{f}_{2}\left(4\times(2,3)\right)$
in a similar way and leave $\tilde{f}_{2}\left(3\times(2,4)\right)$
untouched. After the modification (\ref{eq:mod2}) we construct the corresponding
discrete vector field for $f_{2}$. The Morse complex of $f_{2}$
consists of one critical $0$-cell (vertex $(1,2)$) and two critical
$1$ - cells (edges $3\times(2,4)$ and $1\times(3,4)$). Observe
that there are two different mechanisms responsible for criticality
of these $1$ - cells. The cell $1\times(3,4)$ is critical due to
the definition of trial Morse function $\tilde{f}_{2}$ and $3\times(2,4)$ has been chosen to be critical in order to make
$\tilde{f}_{2}$ the well defined Morse function $f_{2}$. We will
see later that these are in fact the only two ways giving rise to the critical
cells. Notice finally that function $f_{2}$ is in fact a perfect
Morse function and the Morse inequalities for it are equalities.

\noindent We will now consider a more difficult example. The one particle
configuration space, i.e. graph $\Gamma_{2}$ together with the perfect
Morse function and its gradient vector field are shown in figure 8(a)
and 8(b).

\begin{figure}[H]
~~~~~~~~~~~~~~~~~\includegraphics[scale=0.5]{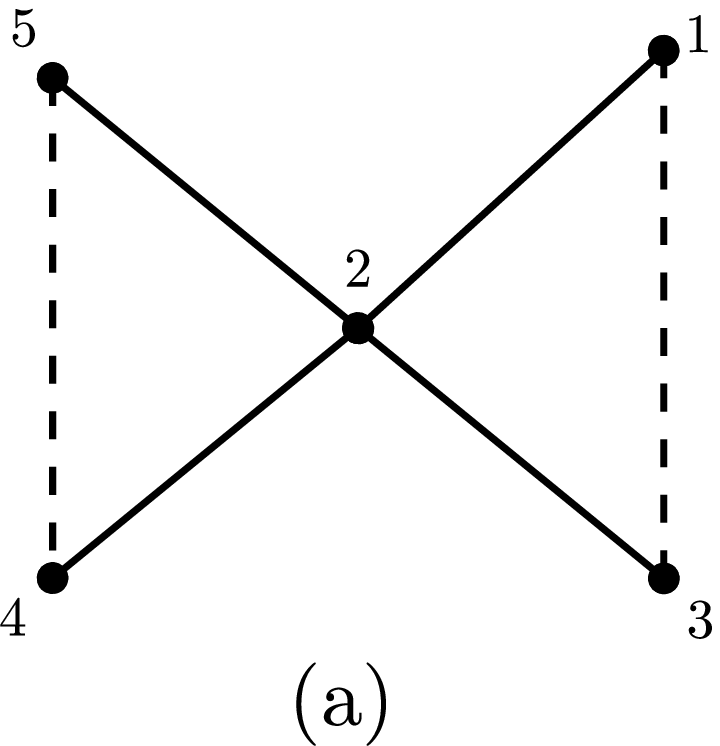}
~~~~~~~~~~~~~~~~~\includegraphics[scale=0.5]{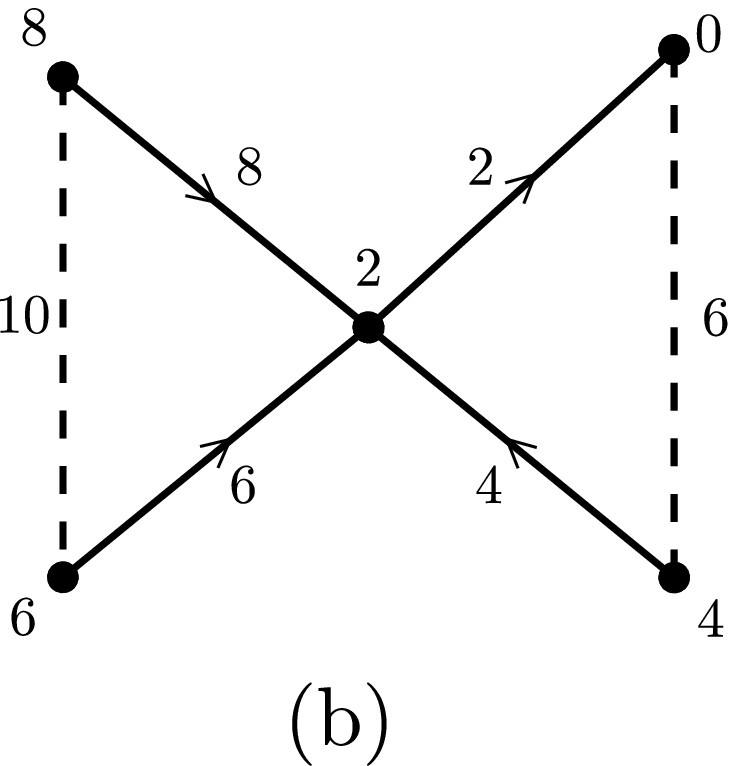}~~~~~~~~

\caption{(a) One particle on bow-tie (b) Perfect discrete Morse function}
\end{figure}

\noindent The construction of two particle configuration space is
a bit more elaborate than in the lasso case and the result is shown
in figure 9(a). Using rules given in (\ref{eq:rules}) we obtain
the trial Morse function $\tilde{f}_{2}$ which is shown in figure
9(b). The critical cells of $\tilde{f}_{2}$ and the cells causing
$\tilde{f}_{2}$ to not be a Morse function are given in table 1.
\begin{table}
\caption{\label{table1}The critical cells of $\tilde{f}_{2}$ and the vertices and edges
causing $\tilde{f}_{2}$ to not be a Morse function.}
\begin{indented}
\item[]\begin{tabular}{|c|c|}
\hline
\multicolumn{2}{|c|}{Critical cells of the trial Morse fuction $\tilde{f}_{2}$}\tabularnewline
\hline
0 - cells & $1\times2$\tabularnewline
\hline
1 - cells & $1\times(4,5)$, $2\times(1,3)$ \tabularnewline
\hline
2 - cells  & $(1,3)\times(4,5)$\tabularnewline
\hline
\end{tabular}
\begin{tabular}{|c|c|c|}
\hline
\multicolumn{3}{|c|}{$\tilde{f}_{2}$ is not Morse function because}\tabularnewline
\hline
vertex  & edges & value\tabularnewline
\hline
$(3,4)$  & $3\times(2,4)$, $4\times(2,3)$ & $\tilde{f}_{2}=10$\tabularnewline
\hline
$(3,5)$ & $5\times(2,3)$, $3\times(2,5)$ & $\tilde{f}_{2}=12$\tabularnewline
\hline
$(4,5)$ & $5\times(2,4)$, $4\times(2,5)$ & $\tilde{f}_{2}=14$\tabularnewline
\hline
\end{tabular}
\end{indented}
\end{table}

In figure 9(b) we have chosen $1$ - cells: $3\times(2,4)$,
$3\times(2,5)$ and $4\times(2,5)$ to be critical, although we should
emphasize that it is one choice out of eight possible ones. We will
now determine the first homology group of the Morse complex $M(f_{2})$
and hence $H_{1}(\mathcal{D}^2(\Gamma_2))$. The Morse complex $M(f_{2})$ is
the sum of $M_{0}(f_{2})$ consisting of one $0$-cell (vertex $1\times2$),
$M_{1}(f_{2})$ which consists of five critical $1$-cells and $M_{2}(f_{2})$
which is one critical $2$-cell $c_{2}=(1,3)\times(4,5)$.
\[
\xymatrix{M_{2}(f_{2})\ar[r]^{\tilde{\partial}_{2}} & M_{1}(f_{2})\ar[r]^{\tilde{\partial}_{1}} & M_{0}(f_{2}).}
\]
The first homology is given by
\begin{eqnarray}
H_{1}(M(f_{2}))=H_{1}(\mathcal{D}^2(\Gamma_2))=\frac{\mbox{Ker}\tilde{\partial}_{1}}{\mbox{Im}\tilde{\partial}_{2}}.
\end{eqnarray}
It is easy to see that $\tilde{\partial}_{1}c_{1}=0$ for any $c_{1}\in M_{1}(f_{2})$
and hence $\mbox{Ker}\tilde{\partial}_{1}=\mathbb{Z}^{5}$. What is
left is to find $\tilde{\partial}c_{2}$ which is a linear combination
of critical $1$-cells from $M_{1}(f_{2})$. According to formula
(\ref{eq:boundary}) we take the boundary of $c_{2}$ in $C_{2}(\Gamma_{2})$
and consider all paths starting from it and ending at the $2$-cells
containing critical $1$-cells (see table 2).
\begin{table}
\caption{\label{table2}The boundary of $c_{2}$.}
\begin{indented}
\item[] \begin{tabular}{|c|c|c|c|}
\hline
boundary of $c_{2}$  &  path & critical $1$ - cells & orientation\tabularnewline
\hline
$1\times(4,5)$ & $\emptyset$ & $1\times(4,5)$ & +\tabularnewline
\hline
$5\times(1,3)$ & %
\begin{tabular}{c}
$5\times(1,3)$, $(2,5)\times(1,3)$, $2\times(1,3)$.\tabularnewline
$5\times(1,3)$, $(2,5)\times(1,3)$, $3\times(2,5)$.\tabularnewline
\end{tabular} & %
\begin{tabular}{c}
$2\times(1,3)$\tabularnewline
$3\times(2,5)$\tabularnewline
\end{tabular} &
\begin{tabular}{c}
-\tabularnewline
-\tabularnewline
\end{tabular}\tabularnewline
\hline
$3\times(4,5)$ & %
\begin{tabular}{c}
$3\times(4,5)$, $(4,5)\times(2,3)$, $2\times(4,5)$, \tabularnewline
$(1,2)\times(4,5)$, $1\times(4,5)$.\tabularnewline
\end{tabular} & $1\times(4,5)$ & -\tabularnewline
\hline
$4\times(1,3)$ & %
\begin{tabular}{c}
$4\times(1,3)$, $(1,3)\times(2,4)$, $2\times(1,3)$.\tabularnewline
$4\times(1,3)$, $(1,3)\times(2,4)$, $3\times(2,4)$.\tabularnewline
\end{tabular} & %
\begin{tabular}{c}
$2\times(1,3)$\tabularnewline
$3\times(2,4)$\tabularnewline
\end{tabular} &
\begin{tabular}{c}
+\tabularnewline
+\tabularnewline
\end{tabular}\tabularnewline
\hline
\end{tabular}
\end{indented}

\end{table}
\noindent Eventually taking into account orientation we get
\begin{eqnarray}
\tilde{\partial}_{2}(c_{2})=1\times(4,5)-3\times(2,5)-2\times(1,3)-1\times(4,5)+\\+3\times(2,4)+2\times(1,3)=-3\times(2,5)+3\times(2,4).
\end{eqnarray}
Hence,
\begin{eqnarray}
H_{1}(\mathcal{D}^2(\Gamma_2))=\frac{\mbox{Ker}\tilde{\partial}_{1}}{\mbox{Im}\tilde{\partial}_{2}}=\mathbb{Z}^{4}.
\end{eqnarray}
The Morse complex $M(f_{2})$ is shown explicitly in figure 9(c).
It is worth mentioning that in this example $f_{2}$ is not a perfect
Morse function.
\begin{figure}[H]
\includegraphics[scale=0.37]{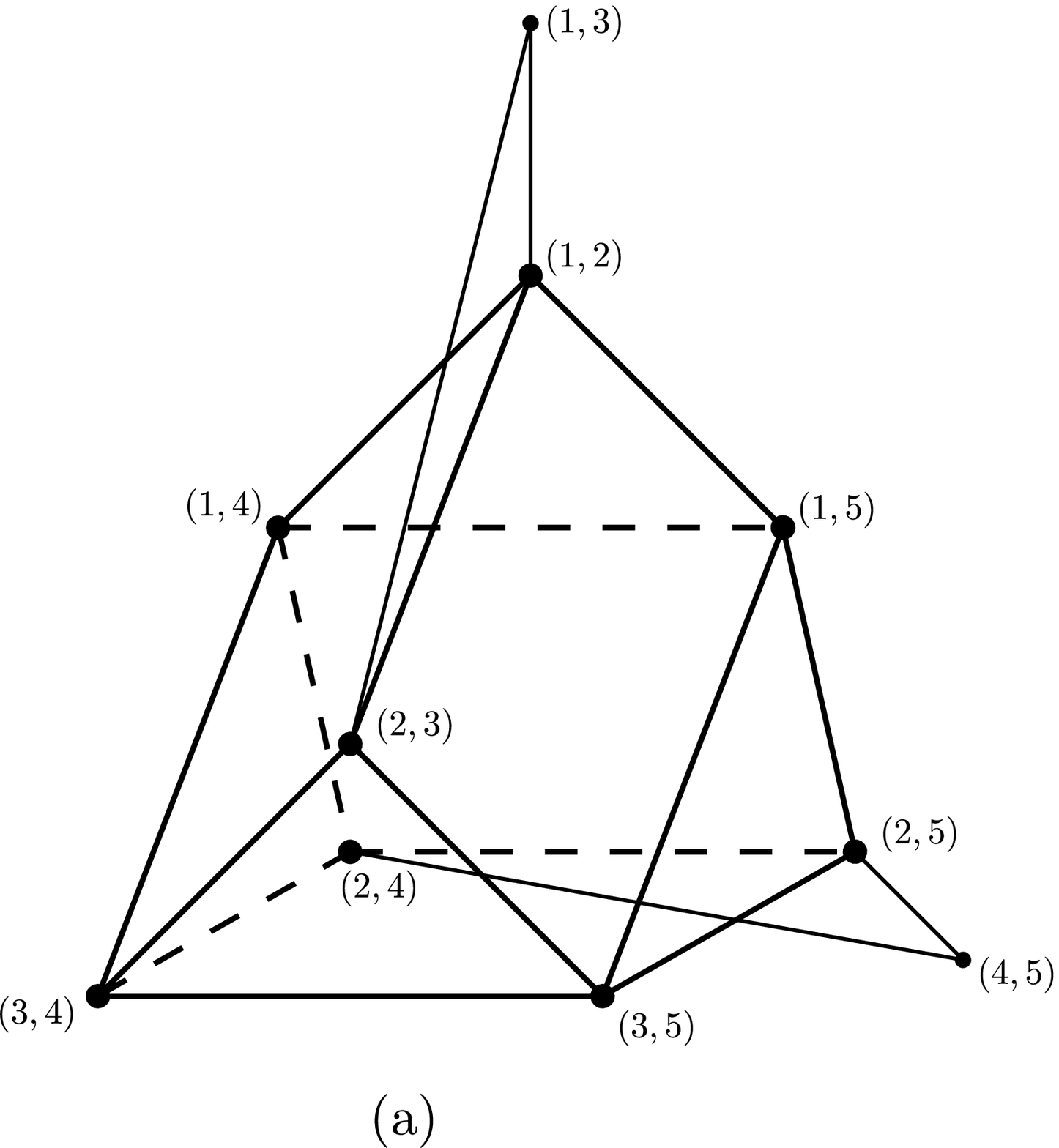}~~~\includegraphics[scale=0.4]{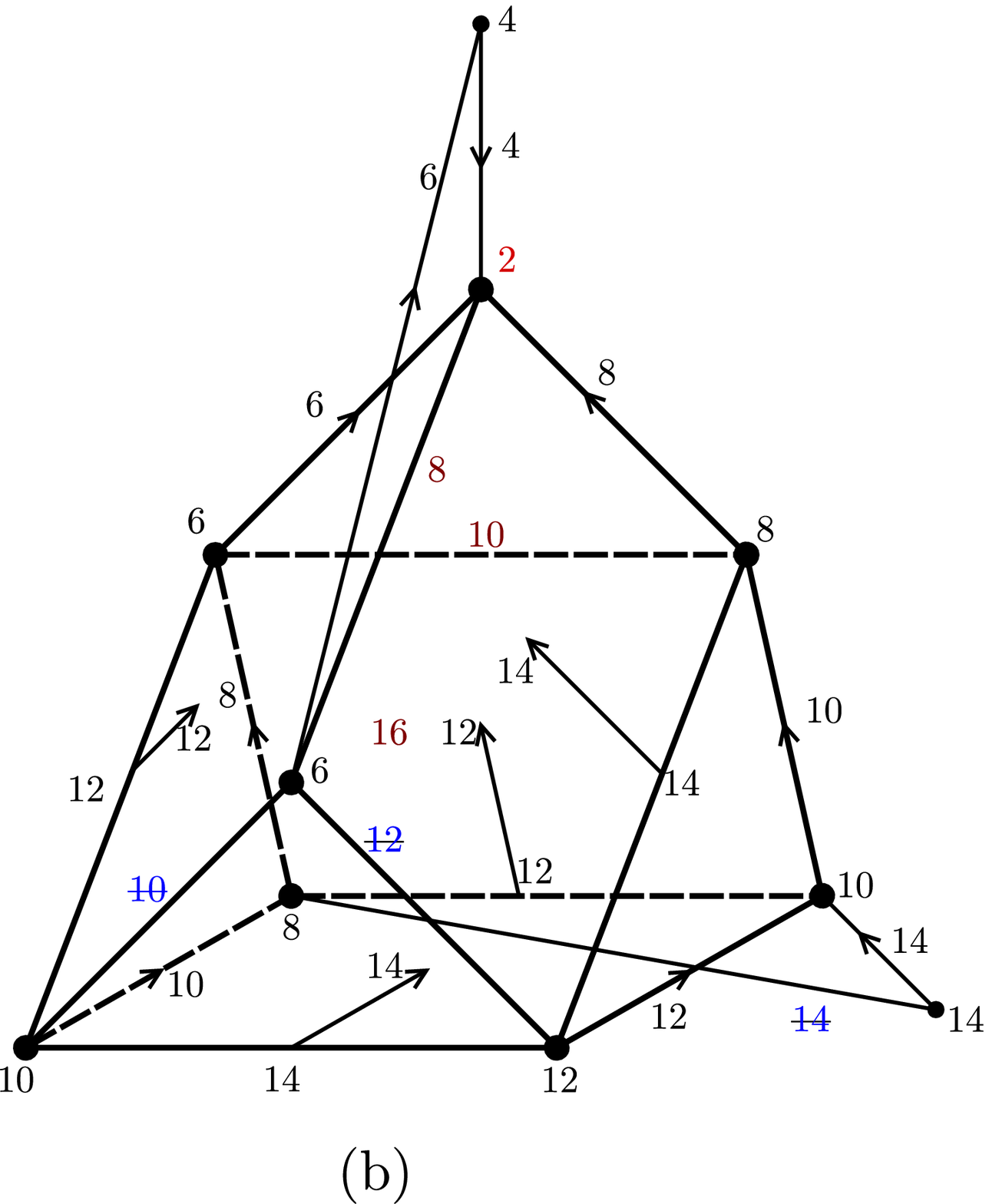}

\medskip{}

~~~~~~~~~~~~~~~~~~~~~~~~~\includegraphics[scale=0.4]{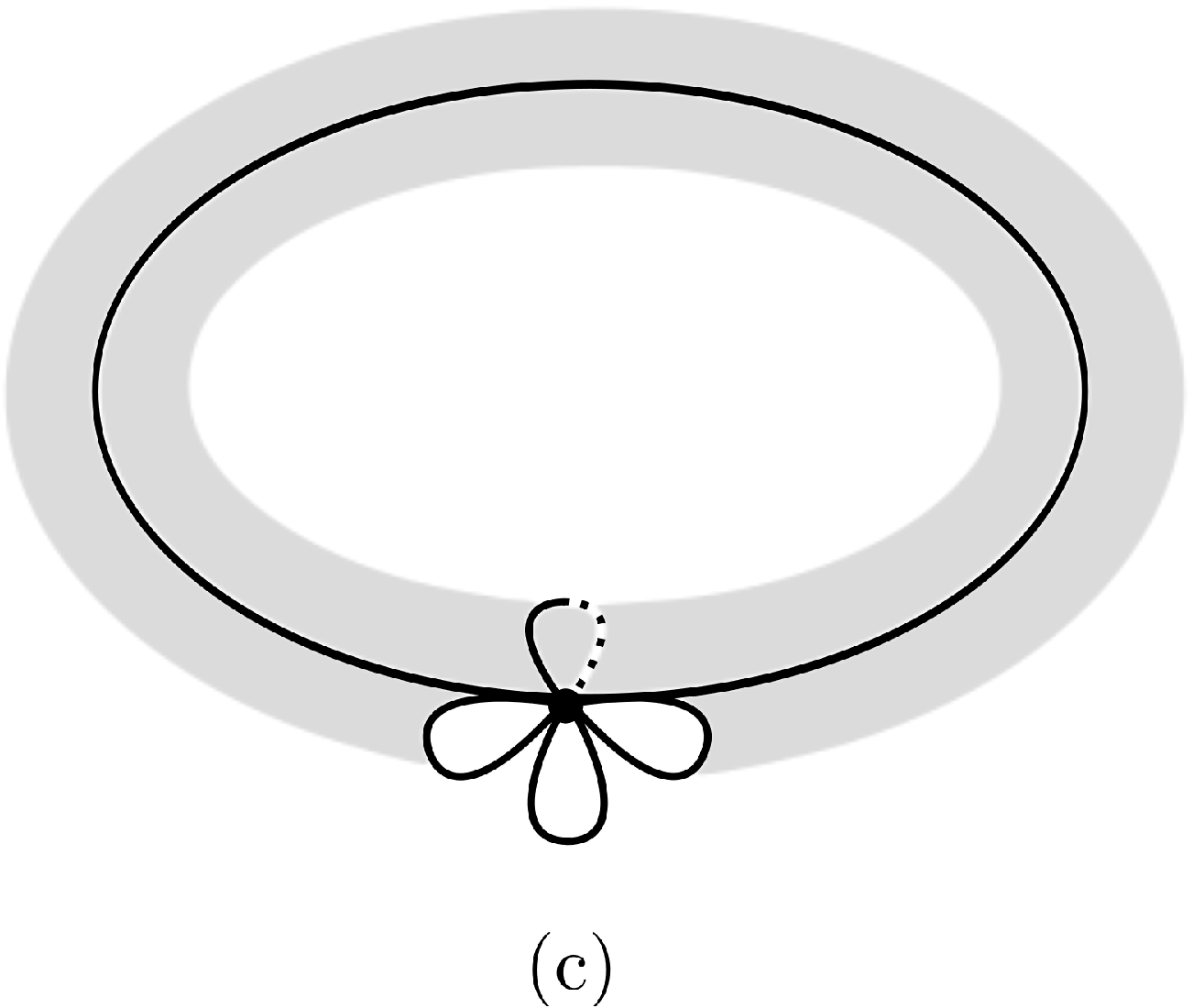}

\caption{(a) Two particles on bow-tie (b) the discrete Morse function and its gradient vector
field, (c) the Morse complex $M(f_{2})$. }
\end{figure}

\section{Discrete Morse theory and topological gauge potentials}

In this section we describe more specifically how the techniques of discrete Morse theory apply to the problem of quantum statistics on graphs. A more general discussion of the model can be found in \cite{JHJKJR}. Here we describe a particular representative example, highlighting the usefulness of discrete Morse theory.

Let $\Gamma$ be a graph shown in figure 7(a). The Hilbert space associated to $\Gamma$ is $\mathcal{H}=\mathbb{C}^4$ and is spanned by vertices of $\Gamma$. The dynamics is given by Schr\"{o}dinger equation where the Hamiltonian $H$ is a hermitian matrix, such that $H_{jk}=0$ if $j$ is not adjacent to $k$ in $\Gamma$. As discussed in \cite{JHJKJR} this corresponds to the so-called tight binding model of one-particle dynamics on $\Gamma$. One can add to the model an additional ingredient, namely whenever particle hops between adjacent vertices of $\Gamma$ the wave function gains an additional phase factor. This can be incorporated to the Hamiltonian by introducing the gauge potential. It is an antisymmetric real matrix $\Omega$ such that each $\Omega_{jk}\in [0,\,2\pi[$ and $\Omega_{jk}=0$ if $j$ is not adjacent to $k$ in $\Gamma$. The modified Hamiltonian is then $H_{jk}^\Omega=H_{jk}e^{i\Omega_{jk}}$. The flux of $\Omega$ through any cycle of $\Gamma$ is the sum of values of $\Omega$ on the directed edges of the cycle. It can be given a physical interpretation of the Aharonov-Bohm phase.

In order to describe in a similar manner the dynamics of two indistinguishable particles on $\Gamma$ we follow the procedure given in \cite{JHJKJR}. The structure of the Hilbert space and the corresponding tight binding Hamiltonian are encoded in $\mathcal{D}^2(\Gamma)$. Namely, we have $\mathcal{H}_2=\mathbb{C}^6$ and is spanned by the vertices of $\mathcal{D}^2(\Gamma)$. The Hamiltonian is given by a hermitian matrix, such that $H_{j,k\rightarrow l}=0$ if $k$ is not adjacent to $l$ in $\Gamma$. The notation $j,k\rightarrow l$ describes two vertices $(j,k)$ and $(j,l)$ connected by an edge in $\mathcal{D}^2(\Gamma)$. The additional assumption which we add in this case stems from the topological structure of $\mathcal{D}^2(\Gamma)$ and is reflected in the condition on the gauge potential. Namely, since the 2-cell $c_2=(1,2)\times(3,4)$ is contractible we require that the flux through its boundary vanishes, i.e.
\begin{equation}\label{Omega}
\Omega(\partial c_2)=\Omega_{1,3\rightarrow 4}+\Omega_{4,1\rightarrow 2}+\Omega_{2,4\rightarrow 3}+\Omega_{3,2\rightarrow 1}=0\, \mathrm{mod}\,2\pi.
\end{equation}
Our goal is to find the parametrization of all gauge potentials satisfying (\ref{Omega}), up to the so-called trivial gauge, i.e. up to addition of $\Omega^\prime$ such that $\Omega^\prime(c)=0\,\mathrm{mod}\,2\pi$, for any cycle $c$. To this end we use discrete Morse theory. We first notice that the edges of $\mathcal{D}^2(\Gamma)$ which are heads of an arrow of the discrete Morse vector field form a tree. Without lose of generality we can put $\Omega_{j,k\rightarrow l}=0$ whenever $j\times(k,l)$ is a head of an arrow. Next, on the edges corresponding to the critical $1$-cells we put arbitrary phases $\Omega_{1,3\rightarrow 4}=\phi$ and $\Omega_{3,2\rightarrow 4}=\psi$. Notice that since $f_2$ is a perfect Morse function these phases are independent. The only remaining edge is $2\times(3,4)$ which is a tail of an arrow. In order to decide what phase should be put on it we follow the gradient path of the discrete Morse vector field which leads to edge $1\times(3,4)$. Hence $\Omega_{2,3\rightarrow 4}=\phi$. The effect of our construction is the topological gauge potential $\Omega$ which is given by two independent parameters (see figure 7(d)) and satisfies (\ref{Omega}). The described reasoning can be \emph{mutatis mutandis} applied to any graph $\Gamma$, albeit the phases on edges corresponding to the critical cells are not independent if $f_2$ is not a perfect Morse function. Finally notice, that in the considered example, the phase  $\phi$ can be interpreted as an Aharonov-Bohm phase and $\psi$ as the exchange phase. The later gives rise to the anyon statistics.

\section{General consideration for two particles\label{sec:General-consideration-for} }
In this section we investigate the first Homology group $H_{1}(C_{2}(\Gamma))$
by means of discrete Morse theory. In section \ref{sec:Main-example} the idea of a trial Morse function
was introduced. Let us recall here that the trial Morse function is
defined in two steps. The first one is to define a perfect Morse function
on $\Gamma$. To this end one chooses the spanning tree $T$ in $\Gamma$.
The vertices of $\Gamma$ are labeled by $1,\,2,\ldots,|V|$ according
to the procedure described in section 4. The perfect Morse function
$f_{1}$ on $\Gamma$ is then given by its value on the vertices and
edges of $\Gamma$, i.e.
\begin{eqnarray}
f_{1}(i)=2i-2,\\
f_{1}((j,k))=\mathrm{max}(f_{1}(j),\, f_{1}(k)),\,\,(j,k)\in T,\\
f_{1}((j,k))=\mathrm{max}(f_{1}(j),\, f_{1}(k))+2,\,\,(j,k)\in\Gamma\setminus T\label{eq:f1-1}
\end{eqnarray}
When $f_{1}$ is specified the trial Morse function on $\mathcal{D}^2(\Gamma)$
is given by the formula
\begin{eqnarray}
\mathrm{0-cells:\,\,\,\,\,\,\,\,\,\,\,\,\,}\,\,\,\,\,\,\,\,\,\,\,\,\tilde{f}_{2}(i\times j) & = & f_{1}(i)+f_{1}(j),\nonumber \\
\mathrm{1-cells:}\,\,\,\,\,\,\,\,\,\,\,\,\tilde{f}_{2}\left(i\times(j,k)\right) & = & f_{1}(i)+f_{1}\left((j,k)\right),\nonumber \\
\mathrm{2-cells:}\,\,\,\tilde{f}_{2}\left((i,j)\times(k,l)\right) & = & f_{1}\left((i,j)\right)+f_{1}\left((k,l)\right).\label{eq:rules-1}
\end{eqnarray}
Let us emphasize that the trial Morse function is typically not a Morse function, i.e., the conditions of definition \ref{Morse-fuction} might not be satisfied. Nevertheless, we will show that it is always possible to modify the function $\tilde{f}_{2}$ and obtain a Morse function $f_2$ out of it. In fact the function $\tilde{f_2}$ is not 'far' from being a Morse function and, as we will see, the number of cells at which it needs fixing is relatively small. In the next paragraphs we localize the obstructions causing $\tilde{f}_2$ to not be a Morse function and explain how to overcome them.

\noindent The cell complex $\mathcal{D}^2(\Gamma)$ consists of $2$, $1$, and $0$-cells which we will denote by $\alpha$, $\beta$ and $\kappa$ respectively. For all these cells we have to verify the conditions of definition \ref{Morse-fuction}. Notice that checking these conditions for any cell involves looking at its higher and lower dimensional neighbours. In case of $2$-cell $\alpha$ we have only the former ones, i.e., the $1$-cells $\beta$ in the boundary of $\alpha$. For the $1$-cell $\beta$ both $2$-cells $\alpha$ and $0$-cells $\kappa$ are present. Finally for the $0$-cell $\kappa$ we have only $1$-cells $\beta$.

Our strategy is the following. We begin with the trial Morse function $\tilde{f}_2$ and go over all $2$-cells checking the conditions of definition \ref{Morse-fuction}. The outcome of this step is a new trial Morse function $\bar{f}_2$ which has no defects on $2$-cells. Next we consider all $1$-cells and verify the conditions of definition \ref{Morse-fuction} for $\bar{f}_2$. It happens that they are satisfied. Finally we go over all $0$-cells. The result of this three-steps procedure is a well defined Morse function $f_2$. Below we present more detailed discussion. The proofs of all statements are in the Appendix.
\begin{enumerate}
    \item \textbf{Step 1} We start with a trial Morse function $\tilde{f}_2$. We notice first that for any edge $e\in T$ there is a unique vertex $v$ in its boundary such that $f_1(e)=f_1(v)$. In other words every vertex $v$, different from $v=1$, specifies exactly one edge $e\in T$ which we will denote by $e(v)$. Next we divide the set of $2$-cells into three disjoint classes. The first one contains $2$-cells $\alpha=e_i\times e_j$, where both $e_i,e_j\notin T$. The second one contains  $2$-cells $\alpha=e_i\times e(v)$, where $e(v)\in T$ and $e_i\notin T$, and the last one contains $2$-cells $\alpha=e(u)\times e(v)$, where both $e(u),e(v) \in T$. Now, since there are no $3$-cells, we have only to check that for each $2$-cell $\alpha$
\begin{eqnarray}\label{alphacond}
    \#\{\beta\subset\alpha\,:\, \tilde{f}_2(\beta)\geq \tilde{f}_2(\alpha)\}\leq1
\end{eqnarray}
The following results are proved in the Appendix
\begin{enumerate}
    \item For the $2$-cells $\alpha=e_i\times e_j$ where both $e_i,e_j\notin T$ the condition (\ref{alphacond}) is satisfied (see fact \ref{fact1}).
    \item For the $2$-cells $\alpha=e_i\times e(v)$ where $e_i\notin T$ and $e(v)\in T$ the condition (\ref{alphacond}) is satisfied (see fact \ref{fact2}).

\item For the $2$-cells $\alpha=e(u)\times\e(v)$ where both $e(u),e(v)\in T$ the condition (\ref{alphacond}) is not satisfied. There are exactly two $1$-cells $\beta_1,\beta_2\subset\alpha$ such that $\tilde{f}_2(\beta_1)=\tilde{f}_2(\alpha)=\tilde{f}_2(\beta_2)$. They are of the form $\beta_1=u\times e(v)$ and $\beta_2=v\times e(u)$. The function $\tilde{f}_2$ can be fixed in two ways (see fact \ref{fact3}). We put $\bar{f}_2(\alpha)=\tilde{f}_2(\alpha)+1$ and either $\bar{f}_2(\beta_1):=\tilde{f}_2(\beta_1)+1$ or $\bar{f}_2(\beta_2):=\tilde{f}_2(\beta_2)+1$.
\end{enumerate}
The result of this step is a new trial Morse function $\bar{f}_2$, which satisfies (\ref{alphacond}).
\item \textbf{Step 2} We divide the set of $1$-cells into two disjoint classes. The first one contains $1$-cells $\beta=v\times e$, where $e\notin T$ and the second one contains $\beta=v\times e(u)$, where $e(u)\in T$. For the $1$-cells within each of this classes we introduce additional division with respect to condition $e(v)\cap e=\emptyset$ (or $e(v)\cap e(u)= \emptyset$). Notice that all $1$-cells $\beta$ which were modified in \textbf{Step 1} belong to the second class and satisfy $e(v)\cap e(u)=\emptyset$. Next we take a trial Morse function $\bar{f}_2$ and go over all $1$-cells $\beta$ checking for each of them if
\begin{eqnarray}\label{betacondition1}
\#\{\alpha\supset\beta\,:\, \bar{f}_2(\alpha)\leq \bar{f}_2(\beta)\}\leq1,\\
\#\{\kappa\subset\beta\,:\, \bar{f}_2\geq \bar{f}_2(\beta)\}\leq1.\label{betacondition2}
\end{eqnarray}
What we find out is
\begin{enumerate}
        \item For the $1$-cells $\beta=v\times e(u)$, where $e(u)\in T$ and $e(v)\cap e(u)\neq\emptyset$ the conditions (\ref{betacondition1}, \ref{betacondition2}) are satisfied (see fact \ref{fact4}). 

    \item For the $1$-cells $\beta=v\times e$, where $e\notin T$ and $e(v)\cap e\neq\emptyset$ the conditions (\ref{betacondition1}, \ref{betacondition2}) are satisfied (see fact \ref{fact5}).
    \item For the $1$-cells $\beta=v\times e(u)$, where $e(u)\in T$ and $e(v)\cap e(u)=\emptyset$ the conditions (\ref{betacondition1}, \ref{betacondition2}) are satisfied (see fact \ref{fact6}).
    \item For the $1$-cells $\beta=v\times e$, where $e\notin T$ and $e(v)\cap e=\emptyset$ the conditions (\ref{betacondition1}, \ref{betacondition2}) are satisfied (see fact \ref{fact7}). 
\end{enumerate}
Summing up the trial Morse function $\bar{f}_2$, obtained in \textbf{Step 1} satisfies both (\ref{alphacond}) and (\ref{betacondition1}), (\ref{betacondition2}). We switch now to the analysis of $0$-cells.
\item \textbf{Step 3} We divide the set of $0$-cells into four disjoint classes in the following way. We denote by $\tau(v)\neq v$ the vertex to which $e(v)$ is adjacent and call it the terminal vertex of $e(v)$. For any $0$-cell $\kappa=v\times u$ we have that either
\begin{enumerate}
    \item $e(v)\cap e(u)\neq\emptyset$ and the terminal vertex $\tau(v)$ of $e(v)$ is equal to $u$.\label{k1}
    \item $e(v)\cap e(u)\neq\emptyset$ and the terminal vertex $\tau(u)$ of $e(u)$ is equal to the terminal vertex $\tau(v)$ of $e(v)$.\label{k2}
    \item $e(v)\cap e(u)=\emptyset$.\label{k3}
    \item $\kappa=1\times u$.\label{k4}
\end{enumerate}
What is left is checking the following condition for any $0$-cell $\kappa$ :
\begin{eqnarray}\label{betacond}
\#\{\beta\supset\kappa\,:\, \bar{f}_2(\beta)\leq \bar{f}_2(\kappa)\}\leq1
\end{eqnarray}
We find out that
\begin{enumerate}
    \item For the $0$-cell $\kappa=u\times v$ belonging to \ref{k1} the condition (\ref{betacond}) is satisfied (see fact \ref{fact8}).
     \item For the $0$-cell $\kappa=u\times v$  belonging to \ref{k2} the condition (\ref{betacond}) is not satisfied. There are exactly two $1$-cells $\beta_1,\beta_2\supset\kappa$ such that $\bar{f}_2(\beta_1)=\bar{f}_2(\kappa)=\bar{f}_2(\beta_2)$. They are of the form $\beta_1=u\times e(v)$ and $\beta_2=v\times e(u)$. The function $\bar{f}_2$ can be fixed in two ways. We put $f_2(\beta_1):=\bar{f}_2(\beta_1)+1$ or $f_2(\beta_2):=\bar{f}_2(\beta_2)+1$ (see fact \ref{fact9}). Moreover, this change does not violate the Morse conditions at any $2$-cell containing $\beta_i$.
     \item For the $0$-cell $\kappa=u\times v$ belonging to \ref{k3} the condition  (\ref{betacond}) is satisfied (see fact \ref{fact10})
     \item For the $0$-cell $\kappa=u\times v$ belonging to \ref{k4} the condition  (\ref{betacond}) is satisfied (see fact \ref{fact11})
\end{enumerate}
\end{enumerate}

\noindent As a result of the above procedure we obtain the Morse function $f_2$. We can now ask the question which cells of $\mathcal{D}^2(\Gamma)$ are critical cells of $f_2$. Careful consideration of the arguments given in facts \ref{fact1}-\ref{fact11} lead to the following conclusions:

\begin{itemize}
\item The $0$-cell is critical if and only if it is $1\times2$
\item The $1$-cell is critical if and only if

\begin{enumerate}
\item It is $v\times e$ where $e\notin T$ and $e(v)\cap e\neq\emptyset$ or $v=1$.
\item Assume that $e(v)\cap e(u)\neq\emptyset$ and the terminal vertex $\tau(u)$ of $e(u)$
is equal to the terminal vertex $\tau(v)$ of $e(v)$. Then either the $1$-cell $v\times e(u)$ or the $1$-cell $u\times e(v)$ is critical, but not both.
\end{enumerate}
\item The $2$-cell is critical if and only if it is $e_{1}\times e_{2}$ where both
$e_{i}\notin T$.
\end{itemize}

These rules are related to those given by Farley and Sabalka in \cite{farley}. As pointed out by an anonymous referee the freedom in choosing noncritical $1$-cells (described in fact 3) and critical $1$-cells (described in fact 9) is also present in Farley and Sabalka's [8] construction.

\section{Summary}
We have presented a description of topological properties of two-particle
graph configuration spaces in terms of discrete Morse theory. Our
approach is through discrete Morse functions, which may be regarded as two-particle potential energies.
We proceeded by introducing a trial Morse function on the full two-particle cell complex, $\mathcal{D}^2(\Gamma)$, which is simply the sum of single-particle potentials on the
one-particle cell complex, $\Gamma$.
We showed that the trial Morse function is close to being a true Morse function
provided that the single-particle potential is a perfect Morse function on $\Gamma$. Moreover, we give an explicit prescription for removing  local defects. The  fixing process is unique modulo the freedom described in facts
\ref{fact3} and \ref{fact9}. The  construction was demonstrated by two examples. A future goal would be to see if these constructions can provide any simplification in understanding of the results of \cite{kiko}.

\section{Acknowledgments}

I would like to thank Jon Keating and Jonathan Robbins for directing me towards the problem of quantum statistics on graphs and fruitful discussions. I am especially in debt to Jonathan Robbins for critical reading of the manuscript and many valuable comments. I would also like to thank the anonymous referees for their invaluable comments and suggestions which improved the final version of the paper. The support by University of Bristol Postgraduate Research Scholarship and Polish MNiSW grant no. N N202 085840 are gratefully acknowledged.

\section*{References}

\section{Appendix}
In this section we give the proofs of the statements made in section \ref{sec:General-consideration-for}. The following notation will be used. We denote by $D_v$ all edges of $\Gamma$ which are adjacent to $v$ and belong to $\Gamma-T$. Similarly by $T_v$ we denote all edges of $\Gamma$ which are adjacent to $v$ and belong to $T$, except one distinguished edge $e(v)\in T$, but not in $T_v$.

\begin{figure}[H]
~~~~~~~~~~~~~~~~~~~~~~~~~~~~~~~~\includegraphics[scale=0.5]{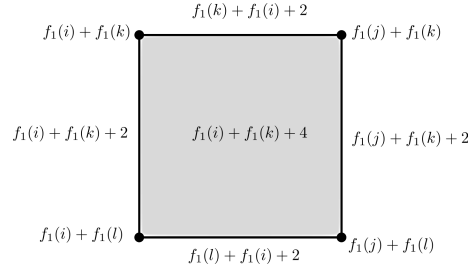}

\caption{The critical cell $e_{1}\times e_{2}$ where both $e_{1}$ and $e_{2}$
do not belong to $T$}\label{figure10}

\end{figure}

\begin{fact}\label{fact1}
Let $\alpha=e_{1}\times e_{2}$ be a $2$-cell such that both $e_{1}$ and $e_{2}$
do not belong to $T$. The condition (\ref{alphacond}) is satisfied and
$\alpha$ is a critical cell.\end{fact}
Proof.
The two cell $e_{1}\times e_{2}$ is shown in the figure \ref{figure10}, where $e_{1}=(i,j)$
and $e_{2}=(k,l)$ and $i>j$, $k>l$. The result follows immediately from this figure.

\begin{figure}[h]


\includegraphics[scale=0.5]{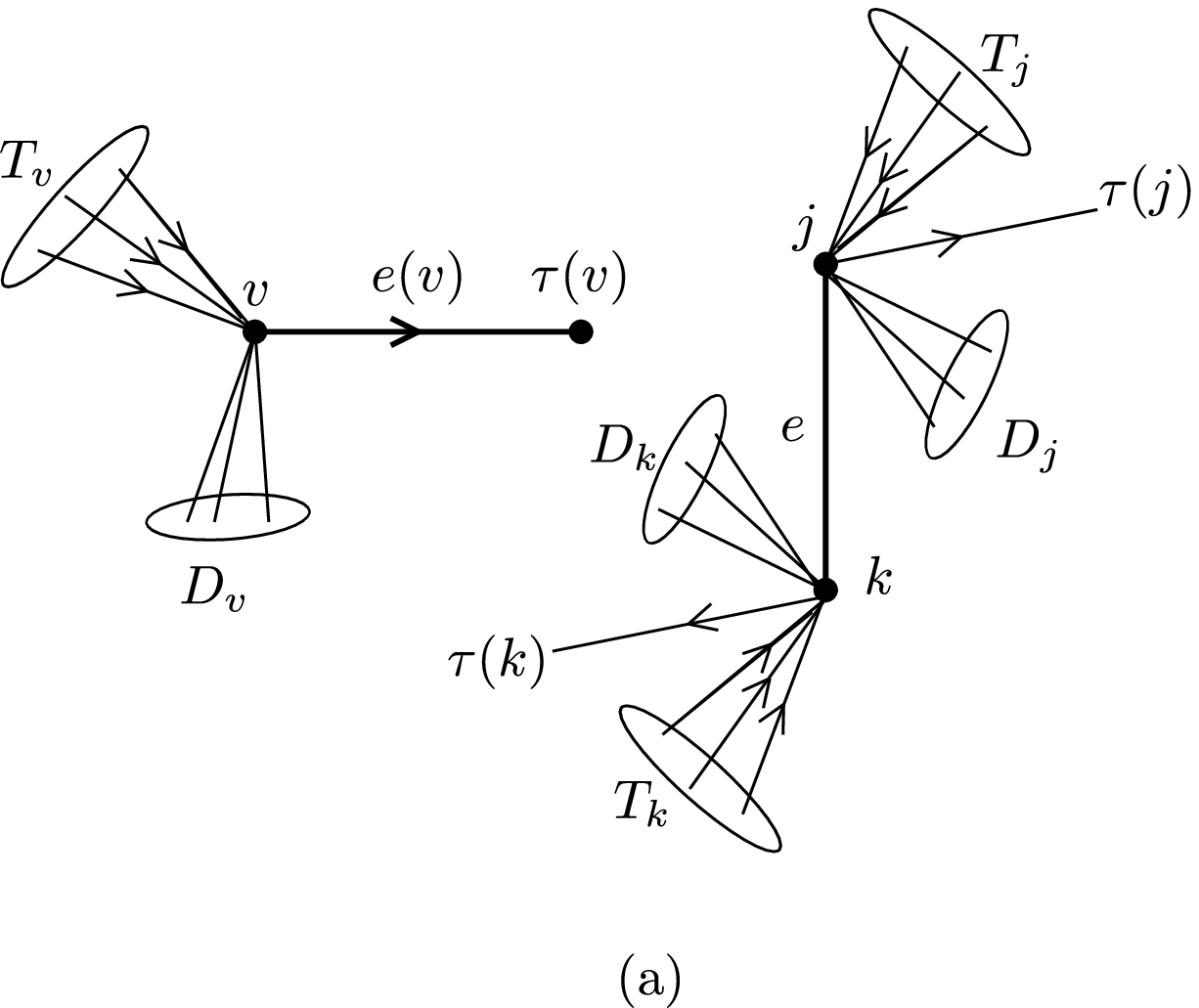}~~\includegraphics[scale=0.5]{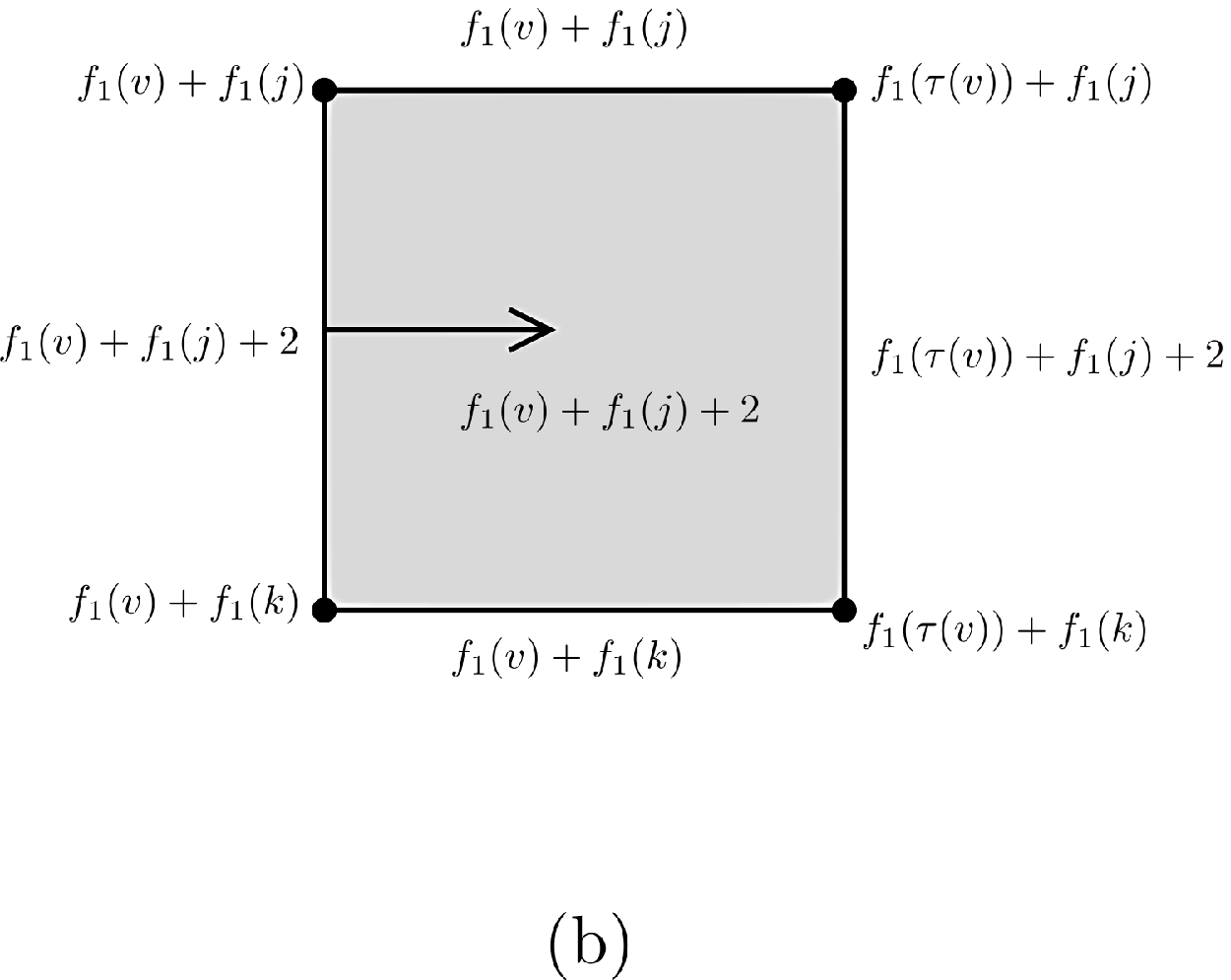}

\caption{(a) $e(v)\cap e=\emptyset$ and $e\notin T$, (b) The
noncritical cells $v\times e$ and $e(v)\times e$. }\label{figure11}
\end{figure}

\begin{fact}\label{fact2}
Let $\alpha=e\times e(v)$ be a $2$-cell, where $e\notin T$ and $e(v)\in T$.
Condition (\ref{alphacond}) is satisfied and
$\alpha$ is a noncritical cell.\end{fact}
Proof.
We of course assume that $e(v)\cap e=\emptyset$. The two cell $\alpha$ is shown on figure \ref{figure11}, where we denoted $e(v)=(v,\tau(v))$ and $e=(j,k)$. The result follows immediately from this figure.

\begin{fact}\label{fact3}
Let $\alpha=e(u)\times\e(v)$ be the $2$-cells, where both $e(u),e(v)\in T$. Condition (\ref{alphacond}) is not satisfied. There are exactly two $1$-cells $\beta_1,\beta_2\subset\alpha$ such that $\tilde{f}_2(\beta_1)=\tilde{f}_2(\alpha)=\tilde{f}_2(\beta_2)$. They are of the form $\beta_1=u\times e(v)$ and $\beta_2=v\times e(u)$. The function $\tilde{f}_2$ can be fixed in two ways. We put $\bar{f}_2(\alpha)=\tilde{f}_2(\alpha)+1$ and either $\bar{f}_2(\beta_1):=\tilde{f}_2(\beta_1)+1$ or $\bar{f}_2(\beta_2):=\tilde{f}_2(\beta_2)+1$. \end{fact}

\noindent Proof.
The $2$-cell $e(v)\times e(u)$ when $e(v)\cap e(u)=\emptyset$
is presented in figure \ref{figure12}(a),(b). The trail Morse function $\tilde{f}_{2}$
requires fixing and two possibilities are shown on figure \ref{figure12}(c),(d).
Notice that in both cases we get a pair of noncritical cells. Namely
the $1$-cell $v\times e(u)$ and $2$-cell $e(v)\times e(u)$ for the situation
presented in figure \ref{figure12}(c) and $1$-cell $u\times e(v)$, $2$-cell $e(v)\times e(u)$
for the situation presented in figure \ref{figure12}(d).

\begin{figure}[H]\label{figure12}
\includegraphics[scale=0.6]{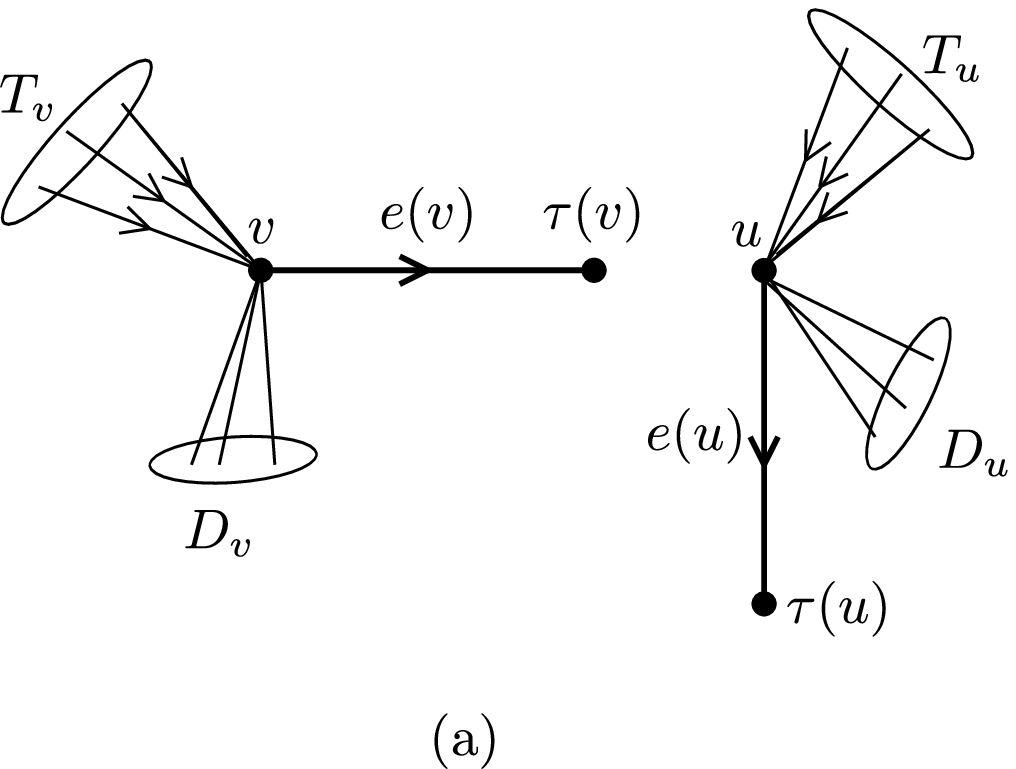}~~~~\includegraphics[scale=0.5]{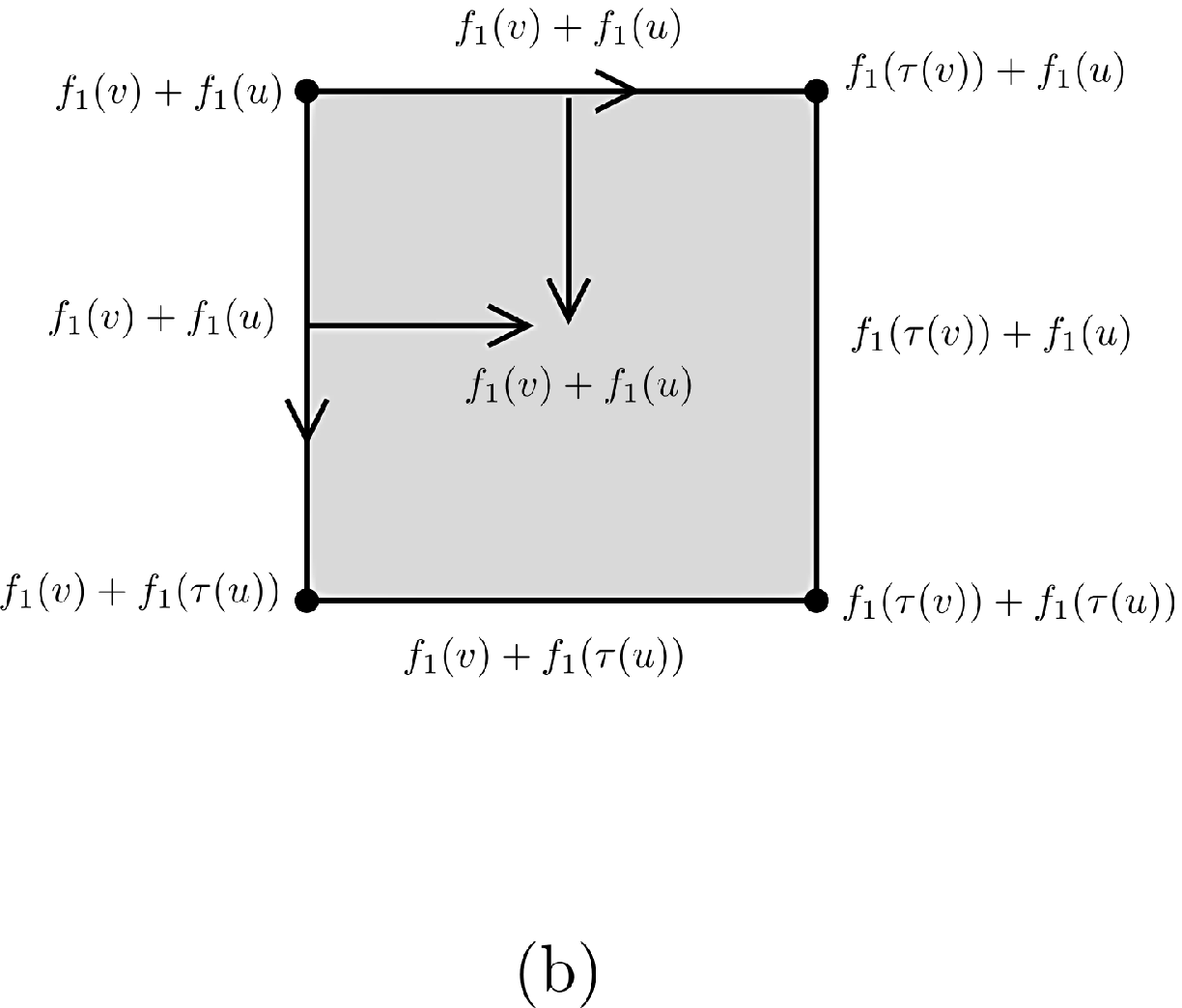}

\bigskip{}

\includegraphics[scale=0.5]{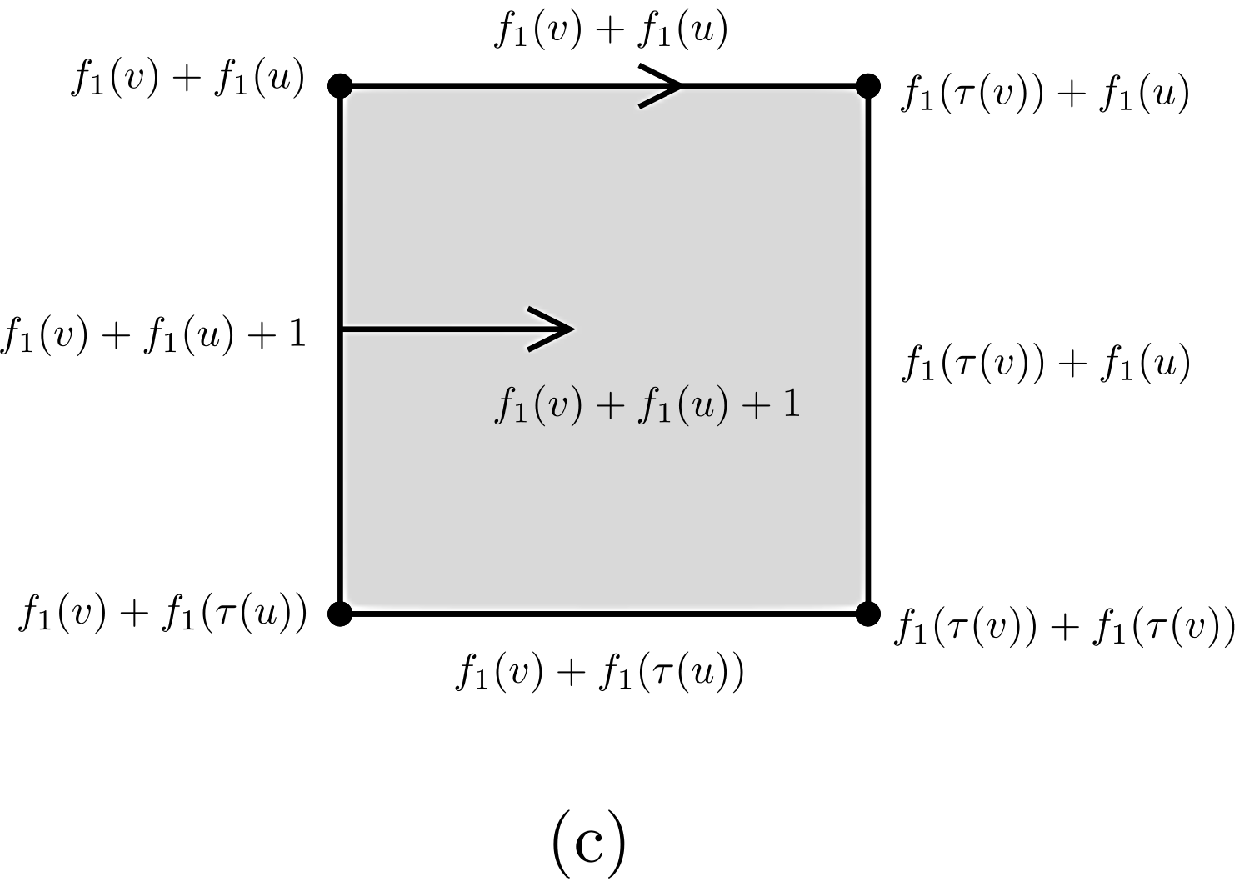}~~\includegraphics[scale=0.5]{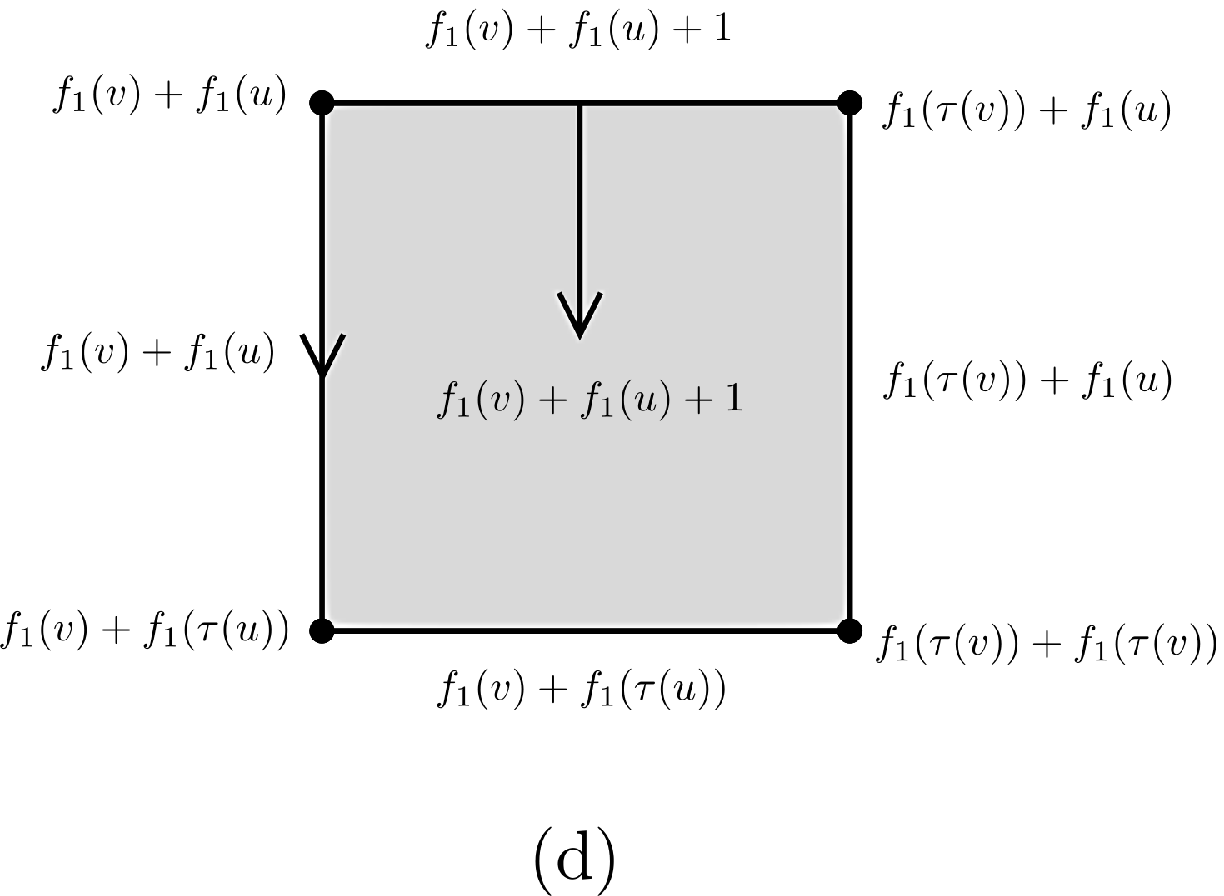}

\caption{(a) Two edges of $T$ with $e(v)\cap e(u)=\emptyset$, (b) The
problem of $2$-cell $e(v)\times e(u)$ (c),(d) two possible fixings
of $\tilde{f}_{2}$ }\label{figure12}
\end{figure}

\begin{fact}\label{fact4}
For the $1$-cells $\beta=v\times e(u)$, where $e(u)\in T$ and $e(v)\cap e(u)\neq\emptyset$ the conditions (\ref{betacondition1}, \ref{betacondition2}) are satisfied.
\end{fact}
\noindent Proof. Let us first calculate $\bar{f}_2(\beta)$. To this end we have to check if $\beta$ was modified in step 1. Notice that every $2$-cell which has $\beta$ in its boundary is one of the following forms:
\begin{enumerate}
    \item $e(v)\times e(u)$\label{1}
    \item $e\times e(u)$ with $e\in D_v$\label{2}
    \item $e\times e(u)$ with $e\in T_v$\label{3}

\end{enumerate}
Case (\ref{1}) is impossible since $e(v)\cap e(u)\neq\emptyset$. For any $2$-cell belonging to (\ref{2}) the value of $\tilde{f}_2$ was not modified on the boundary of $e\times e(u)$ (see fact \ref{fact2}). Finally, for $2$-cells belonging to (\ref{3}) the value of $\tilde{f}_2$ was modified on the boundary of $e\times e(u)$ but not on the cell $\beta$ (see fact 3). Hence $\bar{f}_2(v\times e(u))=\tilde{f}_2(v\times e(u))=f_1(v)+f_1(e(u))=f_1(v)+f_1(u)$. Let us now verify condition (\ref{betacondition2}). The $1$-cell $\beta$ is adjacent to exactly two $0$-cells, namely $v\times u$ and $v\times \tau(u)$. We have $\bar{f}_2(v\times u)=\tilde{f}_2(v\times u)=f_1(v)+f_1(u)$ and  $\bar{f}_2(v\times \tau(u))=\tilde{f}_2(v\times \tau(u))=f_1(v)+f_1(\tau(u))$. Now since $f_1(\tau(u))<f_1(u)$ condition (\ref{betacondition2}) is satisfied. For condition (\ref{betacondition1}) we have only to examine $2$-cells of forms (\ref{2}) and (\ref{3}) (listed above). For $2$-cells that belong to (\ref{2}) we have $f_2(e\times e(u))=f_1(e)+f_1(e(u))>f_1(v)+f_1(u)+2$ and for $2$-cells that belong to (\ref{3}) we have $f_2(e\times e(u))=f_1(e)+f_1(e(u))+1>f_1(v)+f_1(u)+1$. Hence in both cases $\bar{f}_2(e\times e(u))>\bar{f}_2(v\times e(u))$ and condition (\ref{betacondition1}) is satisfied.

\begin{fact}\label{fact5}
For the $1$-cells $\beta=v\times e$, where $e\notin T$ and $e(v)\cap e\neq\emptyset$ conditions (\ref{betacondition1}, \ref{betacondition2}) are satisfied.
\end{fact}
\noindent Proof. Let us first calculate $\bar{f}_2(\beta)$. To this end we have to check if $\beta$ was modified in step 1. Notice that every $2$-cell which has $\beta$ in its boundary is one of the following forms:
\begin{enumerate}
    \item $e(v)\times e$\label{c1}
    \item $e_i\times e$ with $e_i\in D_v$\label{c2}
    \item $e_i\times e$ with $e_i\in T_v$\label{c3}

\end{enumerate}
Case (\ref{c1}) is impossible since $e(v)\cap e\neq\emptyset$. For any $2$-cell belonging to (\ref{c2}) or (\ref{c3}) the value of $\tilde{f}_2$ was not modified on the boundary of $e_i\times e(u)$ (see fact \ref{fact1} and \ref{fact2}). Hence $\bar{f}_2(v\times e)=\tilde{f}_2(v\times e)=f_1(v)+f_1(e)$. Let us now verify condition (\ref{betacondition2}). To this end assume that $e=(j,k)$ with $j>k$. The $1$-cell $\beta$ is adjacent to exactly two $0$-cells, namely $v\times j$ and $v\times k$. We have $\bar{f}_2(v\times j)=\tilde{f}_2(v\times j)=f_1(v)+f_1(j)$ and  $\bar{f}_2(v\times k)=\tilde{f}_2(v\times k)=f_1(v)+f_1(k)$. Now since $f_1(e)=\mathrm{max}(f_1(j),f_1(k))+2$ condition (\ref{betacondition2}) is satisfied. For condition (\ref{betacondition1}) we have only to examine $2$-cells of forms (\ref{c2}) and (\ref{c3}) (listed above). It is easy to see that in both cases  $\bar{f}_2(e_i\times e)>\bar{f}_2(v\times e)$.

\begin{fact}\label{fact6}
For the $1$-cells $\beta=v\times e(u)$, where $e(u)\in T$ and $e(v)\cap e(u)=\emptyset$ conditions (\ref{betacondition1}, \ref{betacondition2}) are satisfied.
\end{fact}

\noindent Proof. Let us first calculate $\bar{f}_2(\beta)$. To this end we have to check if $\beta$ was modified in step 1. Notice that every $2$-cell which has $\beta$ in its boundary is one of the following forms:
\begin{enumerate}
    \item $e(v)\times e(u)$\label{ccc1}
    \item $e\times e(u)$ with $e\in D_v$\label{ccc2}
    \item $e\times e(u)$ with $e\in T_v$\label{ccc3}

\end{enumerate}
For any $2$-cell belonging to (\ref{2}) the value of $\tilde{f}_2$ was not modified on the boundary of $e\times e(u)$ (see fact \ref{fact2}). For the $2$-cells belonging to (\ref{3}) the value of $\tilde{f}_2$ was modified on the boundary of $e\times e(u)$ but not on the cell $\beta$ (see fact 3). Finally for the $2$-cell $e(v)\times e(u)$ the value of $\tilde{f}_2$ was modified on the boundary of $e(v)\times e(u)$ and by fact \ref{fact3} it might be the case that it was modified on $\beta$. Hence $\bar{f}_2(v\times e(u))=\tilde{f}_2(v\times e(u))=f_1(v)+f_1(e(u))=f_1(v)+f_1(u)$ or $\bar{f}_2(v\times e(u))=f_1(v)+f_1(u)+1$. Let us now verify condition (\ref{betacondition2}). The $1$-cell $\beta$ is adjacent to exactly two $0$-cells, namely $v\times u$ and $v\times \tau(u)$. We have $\bar{f}_2(v\times u)=\tilde{f}_2(v\times u)=f_1(v)+f_1(u)$ and  $\bar{f}_2(v\times \tau(u))=\tilde{f}_2(v\times \tau(u))=f_1(v)+f_1(\tau(u))$. Now since $f_1(\tau(u))<f_1(u)$ condition (\ref{betacondition2}) is satisfied. For condition (\ref{betacondition1}) we have to examine $2$-cells from (\ref{ccc1}), (\ref{2}) and (\ref{3}) (listed above). In case when $\bar{f}_2(v\times e(u))=f_1(v)+f_1(u)$ it is easy to see that $\bar{f}_2(e\times e(u))>\bar{f}_2(v\times e(u))$ for $e\in D_v,T_v$ and $\bar{f}_2(e(v)\times e(u))>\bar{f}_2(v\times e(u))$. For $\bar{f}_2(v\times e(u))=f_1(v)+f_1(u)+1$ we still have $\bar{f}_2(e\times e(u))>\bar{f}_2(v\times e(u))$ for $e\in D_v,T_v$ and $\bar{f}_2(e(v)\times e(u))=\bar{f}_2(v\times e(u))$. Hence condition (\ref{betacondition1}) is satisfied in both cases.

\begin{fact}\label{fact7}
For the $1$-cells $\beta=v\times e$, where $e\notin T$ and $e(v)\cap e=\emptyset$ conditions (\ref{betacondition1}, \ref{betacondition2}) are satisfied.
\end{fact}

\noindent Proof. Let us first calculate $\bar{f}_2(\beta)$. To this end we have to check if $\beta$ was modified in step 1. Notice that every $2$-cell which has $\beta$ in its boundary is one of the following forms:
\begin{enumerate}
    \item $e(v)\times e$\label{cc1}
    \item $e_i\times e$ with $e_i\in D_v$\label{cc2}
    \item $e_i\times e$ with $e_i\in T_v$\label{cc3}

\end{enumerate}
For any $2$-cell belonging to (\ref{cc1}), (\ref{cc2}) and (\ref{cc3}) the value of $\tilde{f}_2$ was not modified on the boundary of an appropriate $2$-cell (see fact \ref{fact2} and \ref{fact3}). Hence $\bar{f}_2(v\times e)=\tilde{f}_2(v\times e)=f_1(v)+f_1(e)$. Let us now verify condition (\ref{betacondition2}). To this end assume that $e=(j,k)$ with $j>k$. The $1$-cell $\beta$ is adjacent to exactly two $0$-cells, namely $v\times j$ and $v\times k$. We have $\bar{f}_2(v\times j)=\tilde{f}_2(v\times j)=f_1(v)+f_1(j)$ and  $\bar{f}_2(v\times k)=\tilde{f}_2(v\times k)=f_1(v)+f_1(k)$. Now since $f_1(e)=\mathrm{max}(f_1(j),f_1(k))+2$ condition (\ref{betacondition2}) is satisfied. For condition (\ref{betacondition1}) we have to examine $2$-cells form (\ref{cc1}), (\ref{cc2}) and (\ref{cc3}) (listed above). It is easy to see that  $\bar{f}_2(e_i\times e)>\bar{f}_2(v\times e)$ for $e_i\in D_v,\,T_v$ and $\bar{f}_2(e(v)\times e)=\bar{f}_2(v\times e)$.

\begin{figure}[H]
~~~~~~~~~~~~~~~~~~~~~~~~~~~~~~\includegraphics[scale=0.6]{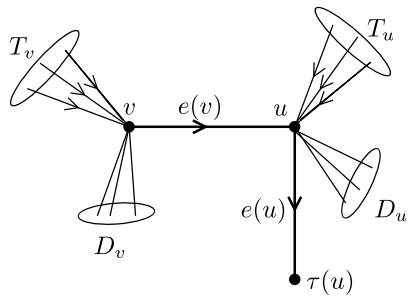}~~~~~~~

\caption{$e(v)\cap e(u)\neq\emptyset$ and $\tau(v)=u$}\label{figure13}

\end{figure}

\begin{fact}\label{fact8}
For the $0$-cell $\kappa=u\times v$ such that $e(v)\cap e(u)\neq\emptyset$ with the terminal vertex $\tau(v)$ of $e(v)$ equal to $u$, condition (\ref{betacond}) is satisfied.
\end{fact}
Proof.
The situation when $e(v)\cap e(u)\neq\emptyset$ and terminal vertex
$\tau(v)$ of $e(v)$ is equal to $u$ is presented in the figure
\ref{figure13}. For the $0$-cell $v\times u$ we have $\bar{f}_2=\tilde{f}_{2}(v\times u)=f_{1}(v)+f_{1}(u)$.
Notice that there is exactly one edge $v\times e(u)$ for which $\bar{f}_{2}\left(v\times e(u))\right)=\bar{f}_{2}(v\times u)$.
The function $\bar{f}_{2}$ on the other edges adjacent to $v\times u$
have a value greater than $\bar{f}_{2}(v\times u)$ and hence $v\times u$
and $v\times e(u)$ constitute a pair of noncritical cells.

\begin{figure}[h]
~~~~\includegraphics[scale=0.6]{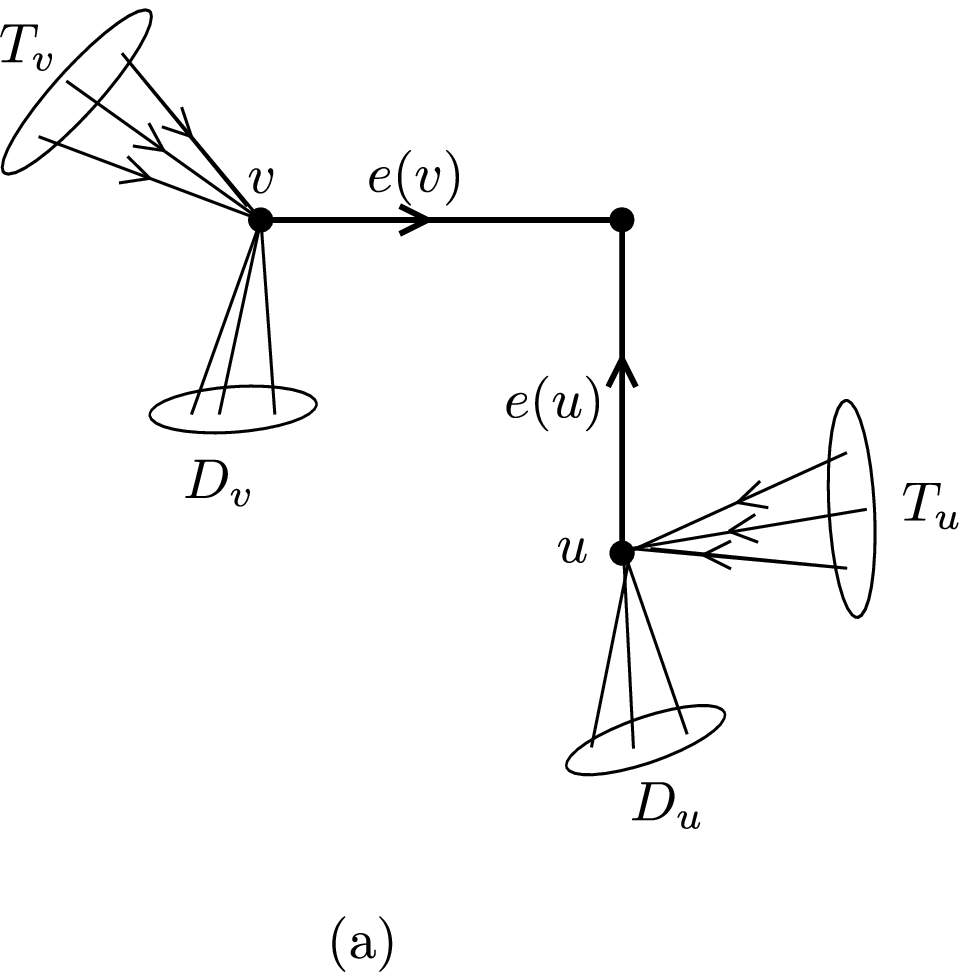}~~~~\includegraphics[scale=0.5]{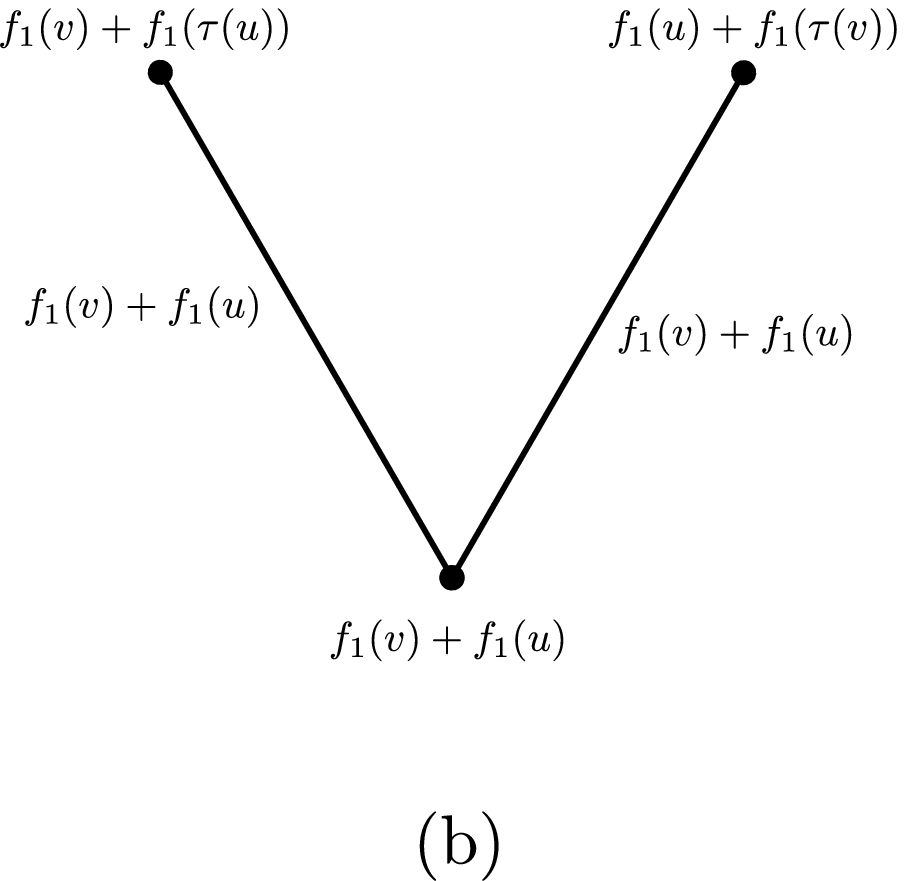}

\bigskip{}

~~~~\includegraphics[scale=0.5]{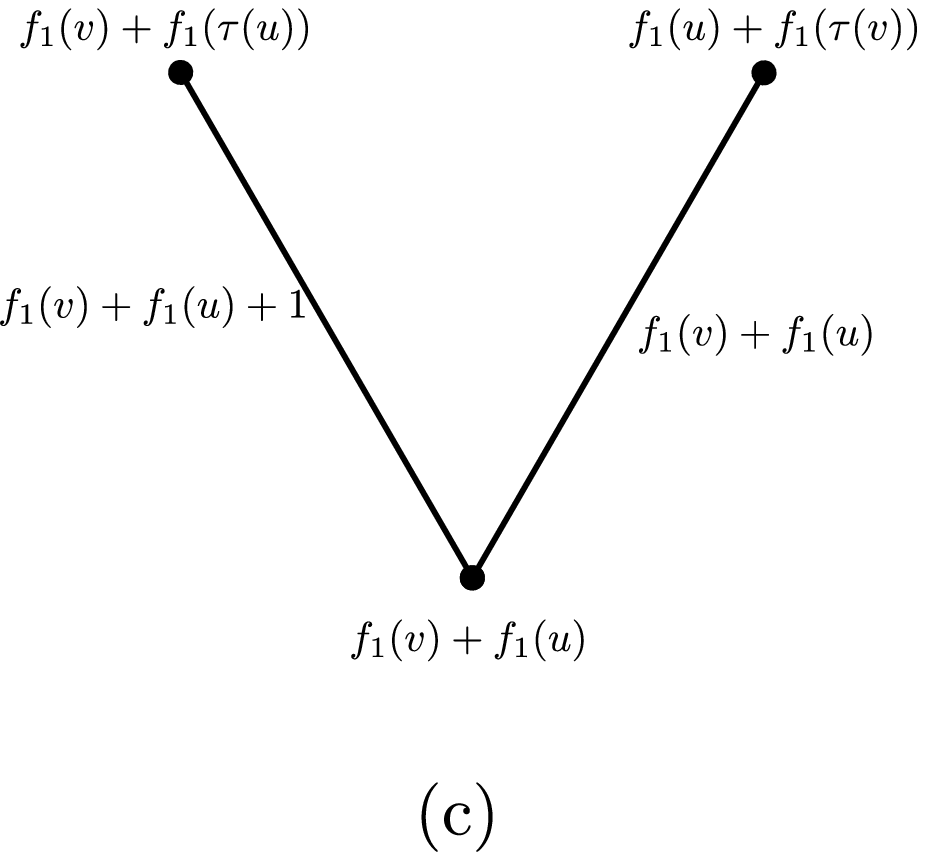}~~~~~\includegraphics[scale=0.5]{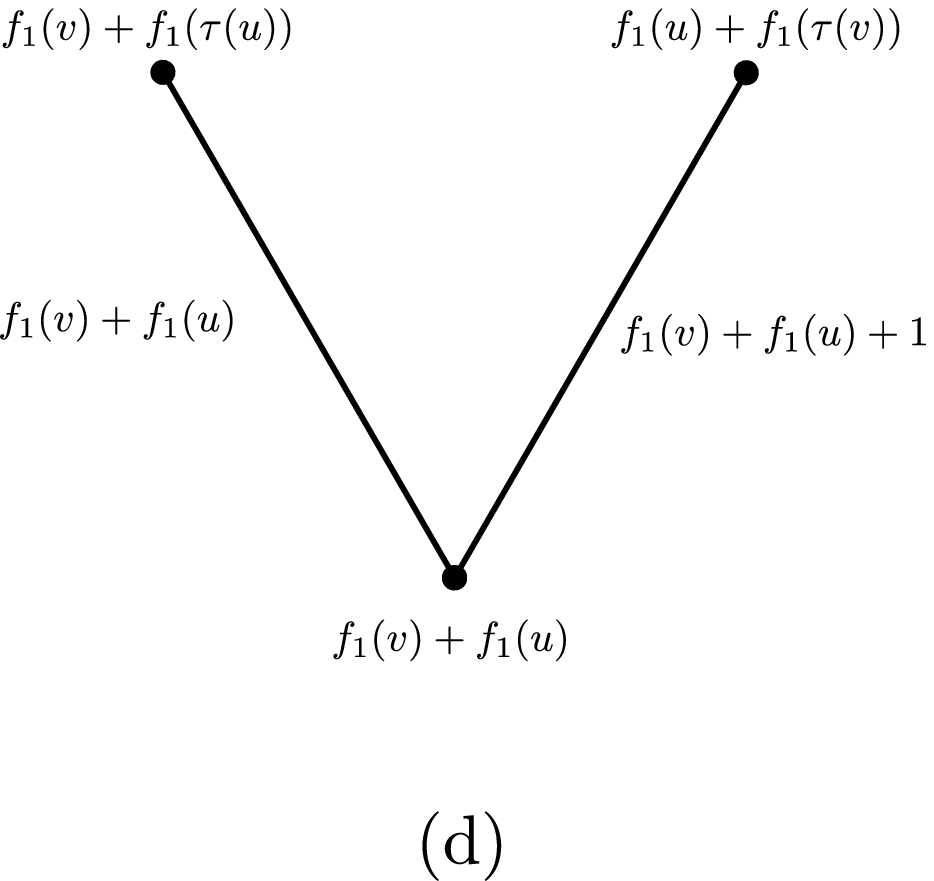}

\caption{(a) Two edges of $T$ with $e(v)\cap e(u)\neq\emptyset$, (b) The problem of $1$-cells $v\times (u,\tau(u))$ and $u\times (v,\tau(v))$ (c),(d) The two possible fixings of $\bar{f}_{2}$}\label{figure14}
\end{figure}

\begin{fact}\label{fact9}
For the $0$-cell $\kappa=u\times v$ such that $e(v)\cap e(u)\neq\emptyset$ with the terminal vertex $\tau(u)$ of $e(u)$ equal to the terminal vertex $\tau(v)$ of $e(v)$ condition (\ref{betacond}) is not satisfied. There are exactly two $1$-cells $\beta_1,\beta_2\supset\kappa$ such that $\bar{f}_2(\beta_1)=\bar{f}_2(\kappa)=\bar{f}_2(\beta_2)$. They are of the form $\beta_1=u\times e(v)$ and $\beta_2=v\times e(u)$. The function $\bar{f}_2$ can be fixed in two ways. We put $f_2(\beta_1):=\bar{f}_2(\beta_1)+1$ or $f_2(\beta_2):=\bar{f}_2(\beta_2)+1$. \end{fact}
\noindent Proof.
The situation when $e(v)\cap e(u)\neq\emptyset$ and terminal vertex
$\tau(u)$ of $e(u)$ is equal to terminal vertex $\tau(v)$ of $e(v)$
is presented in the figure \ref{figure14}(a),(b). For the $0$-cell $v\times u$ we
have $\bar{f}_2(v\times u)=f_{1}(v)+f_{1}(u)$. There are two
edges $v\times e(u)$ and $u\times e(v)$ such that $\bar{f}_{2}(v\times e(u))=\bar{f}_{2}(v\times u)=\bar{f}_{2}(u\times e(v))$.
It is easy to see that the value of $\bar{f}_{2}$ on the other edges adjacent to $v\times u$
is greater than $\bar{f}_{2}(v\times u)$. So the function $\bar{f}_{2}$
does not satisfy condition (\ref{betacond}) and there are two possibilities \ref{figure14}(c),(d) to fix this problem. Either we put
$\bar{f}_{2}(v\times e(u))=\bar{f}_{2}(v\times u)+1$ or
$\bar{f}_{2}(u\times e(v))=\bar{f}_{2}(v\times u)+1$. They
both yield that the vertex $v\times u$ is non-critical. Notice finally that by the definitions of $f_1$ and $\tilde{f}_2$, increasing the value of $\bar{f}_2(\beta_i)$ by one does not influence $2$-cells containing $\beta_i$ in their boundary.

\begin{fact}\label{fact10}
For the $0$-cell $\kappa=u\times v$ such that $e(v)\cap e(u)=\emptyset$ condition  (\ref{betacond}) is satisfied.
\end{fact}
\noindent Proof. This is a direct consequence of the modification made for the $2$-cell $\alpha=e(v)\times e(u)$ in step 1. Moreover, $\kappa$ is noncritical.

\begin{fact}\label{fact11}
For the $0$-cell $\kappa=1\times u$ condition  (\ref{betacond}) is satisfied.
\end{fact}
\noindent Proof.  For the $0$-cell $1\times u$ we have $\bar{f}_2=\tilde{f}_{2}(v\times u)=f_{1}(u)$.
Notice that there is exactly one edge $1\times e(u)$ for which $\bar{f}_{2}\left(1\times e(u)\right)=\bar{f}_{2}(1\times u)$.
The function $\bar{f}_{2}$ on the other edges adjacent to $1\times u$
have a value greater than $\bar{f}_{2}(1\times u)$. Hence if $u\neq 2$ the $0$-cell  $1\times u$
and the $1$-cell $1\times e(u)$ constitute a pair of noncritical cells. Otherwise $\kappa$ is a critical $0$-cell.


\begin{thebibliography}{9}
\bibitem{Souriau70}Souriau, J M 1970 Structure des systèmes dynamiques, Dunod, Paris.

\bibitem{leinass}Leinaas J M, Myrheim J 1977 On the theory of identical
particles. Nuovo Cim.37B, 1\textendash{}23.

\bibitem{wilczek}Wilczek, F (ed.) 1990 Fractional statistics and anyon superconductivity. Singapore, Singapore:
World Scientific.

\bibitem{Dowker85}Dowker, J S 1985 Remarks on non-standard statistics J. Phys. A: Math. Gen. 18 3521

\bibitem{JHJKJR}Harrison J M, Keating  J P and Robbins J M 2011  Quantum
statistics on graphs Proc. R. Soc. A 8 January  vol. 467 no. 2125
212-233

\bibitem{BE} Balachandran A P, Ercolessi E 1992 Statistics on networks. Int. J. Mod. Phys. A 7, 4633–4654.

\bibitem{Abrams}Abrams A 2000 Configuration spaces and braid groups of
graphs. Ph.D. thesis, UC Berkley.

\bibitem{PS09}Prue P, Scrimshaw T 2009 Abrams's stable equivalence for graph braid groups. arXiv:0909.5511

\bibitem{farley}Farley D,  Sabalka L 2005 Discrete Morse theory and graph
braid groups Algebr. Geom. Topol. 5  1075-1109

\bibitem{kiko}Ko K H, Park H W 2011 Characteristics of graph braid groups. arXiv:1101.2648

\bibitem{forman}Forman R 1998 Morse Theory for Cell Complexes Advances
in Mathematics 134, 90145

\bibitem{Fox}Fox R H, Neuwirth L 1962 The braid groups, Math. Scand.
10, 119-126.

\bibitem{Hatcher}Hatcher A 2002 Algebraic Topology, Cambridge University
Press

\bibitem{Ghrist}Ghrist R 2007 Configuration spaces, braids and robotics.
Notes from the IMS Program on Braids, Singapore

\bibitem{milnor}Milnor J 1963 Classical Morse Theory, Princeton University
Press
\bibitem{Ayala11} Ayala R, Fernandez-Ternero D, Vilches J A 2011 Perfect discrete Morse functions on 2-complexes. Pattern Recognition Letters, Available online 10.1016/j.patrec.2011.08.011.
\end{thebibliography}
\end{document}